\RequirePackage{ifpdf}
\documentclass[12pt]{JHEP3}
\pdfoutput=1

\usepackage{amsmath,epsfig}
\usepackage{amssymb,amsfonts}
\usepackage{latexsym}
\usepackage{bbm}
\usepackage[latin1]{inputenc}
\usepackage{subeqnarray}
\usepackage{xcolor}
\let\normalcolor\relax
\usepackage{graphicx}

\usepackage{longtable}

\relax

\def\be{\begin{equation}}
\def\ee{\end{equation}}
\def\bea{\begin{eqnarray}}
\def\eea{\end{eqnarray}}
\newcommand\fverb{\setbox\pippobox=\hbox\bgroup\verb}
\newcommand\fverbdo{\egroup\medskip\noindent%
                        \fbox{\unhbox\pippobox}\ }
\newcommand\fverbit{\egroup\item[\fbox{\unhbox\pippobox}]}

\def\abs#1{\mid \! #1 \! \mid}

\newcommand{\bear}{\begin{eqnarray}}

\newcommand{\eear}{\end{eqnarray}}

\newcommand{\bsea}{\begin{subeqnarray}}
\newcommand{\esea}{\end{subeqnarray}}
\newbox\pippobox

\newcommand{\ba}{\begin{eqnarray}}
\newcommand{\ea}{\end{eqnarray}}

\def\6{\partial}

\def\a{\alpha}

\def\sq
\def\a{\alpha}

\title{Towards a Holographic Bose-Hubbard Model}

\author{{\large Mitsutoshi Fujita$^{1}$, Sarah M. Harrison$^{2}$, Andreas Karch$^3$, Ren\'e Meyer$^4$, and Natalie M. Paquette$^{5}$}\\
$^1$
\href{http://www.yukawa.kyoto-u.ac.jp/english/index.html}{Yukawa Institute for Theoretical Physics}, Kyoto University, Kyoto 606-8502, Japan. \href{mailto:mitsutoshi.fujita@yukawa.kyoto-u.ac.jp}{mitsutoshi.fujita@yukawa.kyoto-u.ac.jp}\\
$^2$
\href{http://hetg.physics.harvard.edu}{Center for the Fundamental Laws of Nature}, Harvard University,
Cambridge, MA 02138, USA. \href{mailto:sarharr@stanford.edu}{sarharr@stanford.edu}\\
$^3$
\href{http://www.phys.washington.edu}{Department of Physics, University of Washington},
 Seattle, WA 98195-1560, USA. \href{mailto:akarch@uw.edu}{akarch@uw.edu}\\
$^4$
\href{http://www.ipmu.jp/}{Kavli Institute for the Physics and Mathematics of the Universe (WPI)}, Todai Institutes for Advanced Study, The University of Tokyo, Kashiwa, Chiba 277-8568, Japan. \href{mailto:rene.meyer@ipmu.jp}{rene.meyer@ipmu.jp}\\
$^5$
\href{http://web.stanford.edu/group/sitp/}{SITP}, \href{https://physics.stanford.edu/}{Department of Physics} and \href{http://www.slac.stanford.edu/th/th.html}{Theory Group, SLAC}, Stanford University, Stanford, CA 94305, USA. \href{mailto:npaquett@stanford.edu}{npaquett@stanford.edu}
}

\preprint{arXiv:1411.7899 [hep-th] \hspace{5cm} 
IPMU14-0283, YITP-14-87}

\abstract{We present a holographic construction of the large-$N$ Bose-Hubbard
model. The model is based on Maxwell fields coupled to charged scalar
fields on the $AdS_2$ hard wall. We realize the lobe-shaped phase
structure of the Bose-Hubbard model and find that the model admits Mott
insulator ground states in the limit of large Coulomb repulsion. In the
Mott insulator phases, the bosons are localized on each site. At zero
hopping we find that the transitions between Mott insulating phases with
different fillings correspond to first order level-crossing phase
transitions. At finite hopping we find a holographic phase transition
between the Mott phase and a non-homogeneous phase. We then analyze the
perturbations of fields around both the Mott insulator phase and
inhomogeneous phase. We find almost zero modes in the non-homogeneous
phase. 
}

\begin{document}

%%%%%%%%%%%%%%%%%%%%%%%%%%%%%%%%%%%%%%%%%%%%%%
%%%%%%%%%%%%%%%%%%%%%%%%%%%%%%%%%%%%%%%%%%%%%%
%%%%%%%%%%%%%%%%%%%%%%%%%%%%%%%%%%%%%%%%%%%%%%

\section{Introduction \& Summary}

Strongly-coupled many-body systems such as the QGP~\cite{DeWolfe:2013cua} have been studied using the gauge/gravity correspondence~\cite{Maldacena:1997re}, which includes useful string theory embeddings of systems which share many qualitative properties of QCD at both zero and finite temperature/density. Following these advances, holographic models of condensed matter phenomena have been constructed, such as an analog of the particle-vortex duality~\cite{Herzog:2007ij,Myers:2010pk}, superfluid-insulator transitions~\cite{Hartnoll07}, superconductivity~\cite{Gubser:2008px,Hartnoll:2008vx,Hartnoll:2008kx}, and the field theory with Lifshitz symmetry~\cite{Kachru:2008yh} as well as hyperscaling violation \cite{Charmousis:2010zz}. However, many of these models assume translation invariance. In the presence of finite charge density, the DC conductivity becomes infinite unlike in real materials which have a Drude peak characterized by a finite DC conductivity.
Recently, holographic models have also been applied to systems in which translation invariance is broken spontaneously, by a periodic function of the chemical potential~\cite{Horowitz:2012ky,Ling:2014saa} or by that of scalars~\cite{Donos:2013eha,Donos:2014uba,Iizuka:2012iv,Iizuka:2012pn} to form a lattice. Interestingly,  physics without translation invariance can also be captured by introducing a lattice of impurities dual to probe $D5(\overline{D5})$ branes wrapped on an asymptotic $AdS_2$ in the $AdS_5$ or $AdS_5$ black hole background. In Ref.~\cite{Kachru:2009xf,Kachru:2010dk}, a dimerization transition on a lattice, which changes the structure of the Fermi surface, was analyzed by coupling defect fermions on D5-$\overline{D5}$ branes with itinerant fermions. Many interesting features of the lattice formulation can already be captured in the probe limit~\cite{Karch:2003nh,Karch:2007pd}. These simple holographic probe models however do not easily allow charge transport of fermions on the probe brane.

In this paper, we are interested in and focus on an approximate (bottom-up) model of interacting bosons on a lattice, including charge transport: the Bose-Hubbard model. The Bose-Hubbard model is e.g. realized in ultra-cold atomic experiments using $^{87}$Rb trapped in an optical lattice~\cite{Markus12}, as well as in Helium atoms moving on substrates. The Bose-Hubbard Hamiltonian is given by
\ba\label{BHHamiltonian}
H=-\sum_{\langle ij\rangle}(t_{hop} \, b_i^{\dagger}b_j+c.c.)+\dfrac{U}{2}\sum_i n_i(n_i-1) -\mu_b \sum_i n_i,
\ea
where $t_{hop}$ is the hopping parameter giving the mobility of the bosons between neighboring sites and $U$ is the remnant of the repulsive Coulomb interaction\,\footnote{Long-range interactions can also be included by generalizing \eqref{BHHamiltonian} beyond on-site and nearest-neighbor interactions, c.f. e.g. \cite{Fisher:1989zza}.} between bosons on a single site. $b_j$ and $b_j^\dagger$ are, respectively, annihilation and creation operators for the bosons at site $j$. $n_i=b^\dagger_i b_i$ (no summation over $i$) is the boson density on site $i$, and $\mu_b$ is the chemical potential. In the Bose-Hubbard model without disorder, there exist only two phases, namely, the Mott insulator phase and the superfluid phase. In the Mott insulator phase, bosons are localized on the lattice due to the repulsive interactions. They do not form a coherent state. In the coherent superfluid phase, bosons are delocalized on the lattice and an off-diagonal long-range order, i.e. long-range correlations $\langle b_{i}^{\dagger}b_{j} \rangle$, exists in the superfluid phase. It is known that this condensate becomes of the same order as the particle density. Our holographic model will show the same pattern. When $U/t_{hop}$ is large, bosons are localized and the ground state is in the Mott insulator phase, while when $U/t_{hop}$ is small, bosons acquire kinetic energy derived from the non-zero hopping parameter $t_{hop}$ and are delocalized.

In the rest of the introduction, we will first review the salient details of the mean-field approach of  \cite{Fisher:1989zza} to the Bose-Hubbard model, and the physics of the zero-temperature phase diagram (\S \ref{sec:rev}). Then in \S \ref{sec:holographic} we will introduce our holographic model and compare our results to the phase diagram of \cite{Fisher:1989zza}.

\subsection{Review of mean-field approach to the Bose-Hubbard model}\label{sec:rev}
In this section we review the field theoretic approach of \cite{Fisher:1989zza} to the phase diagram of the Bose-Hubbard model in the $\mu_b-t_{hop}$ plane at zero temperature. We will then attempt to construct a holographic model which exhibits the same physics. First let's consider the Hamiltonian (\ref{BHHamiltonian}) with $t_{hop}=0$. This describes the phase structure of a lattice of bosons with no hopping as a function of chemical potential. The ground state will be the one which minimizes the potential energy,
\be
\epsilon_i(n)=-\dfrac{U}{2} n_i(n_i-1) +\mu_b  n_i,
\ee
at each site. Taking $U >0$, the solution is that $n$ bosons occupy each site for $\mu_b$ in the range $n-1<\mu_b/U<n$. If $\mu_b<0$, then $n_i=0$ at all sites. If $\mu_b/U$ is exactly a positive integer, $m$, then $\epsilon(m)=\epsilon(m+1)$, and  configurations with $m$ and $m+1$ bosons at a single site are degenerate, leading to $2^N$ possible degenerate ground states for a system with $N$ sites. We see that as we increase $\mu_b/U$ with $t_{hop}=0$, the system undergoes a series of level-crossing phase transitions at each integer value of $\mu_b/U$ where the density of bosons on each site jumps from $n_i=n$ to $n_i=n+1$.

Now let's consider what happens when one turns on a small hopping $t_{hop}>0$ when in a phase with $n$ bosons per site, with $\mu_b/U=n-1/2+\alpha$ for $-1/2<\alpha<1/2$. The energy required to add a particle to the system is $\delta E_P\sim (1/2-\alpha)U$, and to remove a particle (create a hole) is $\delta E_h\sim (1/2+\alpha)U$. For $t_{hop}\ll U$, the kinetic energy gained from allowing an extra particle or hole to hop around the lattice is not large enough to overcome the cost in potential energy of removing a particle or adding a particle. Therefore, for each value of the density $n$, there is some finite region of fixed density with an energy gap for the creation of particle-hole excitations, where the size of the gap decreases with increasing $t_{hop}$. Furthermore, this constant-density state is a Mott insulating phase as the fixed density implies it is incompressible.

\begin{figure}[htbp]
  \begin{center}
   \includegraphics[height=6cm]{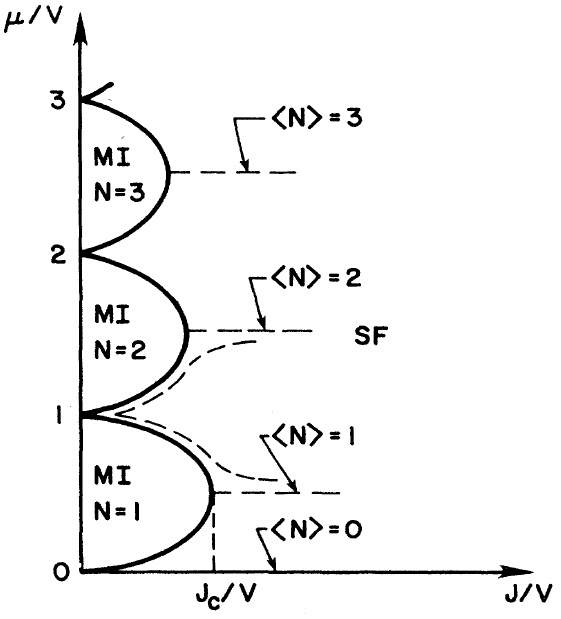}
    \includegraphics[height=5.5cm]{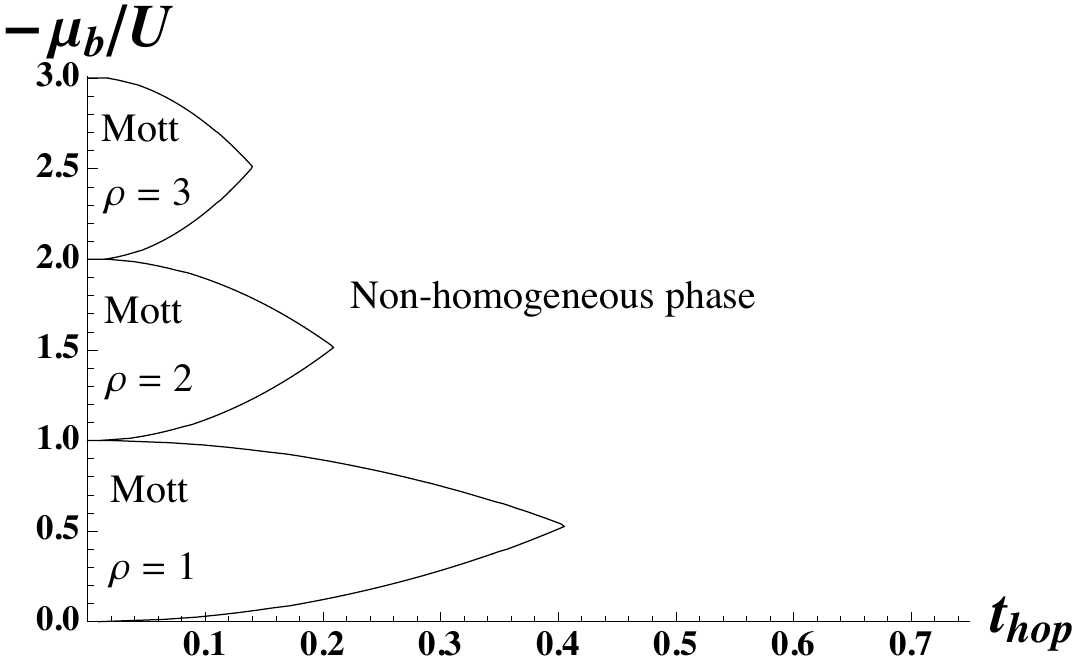}
  \caption{\textbf{Left panel:} This figure, taken from \cite{Fisher:1989zza}, illustrates the zero-temperature phase diagram as a function of chemical potential and hopping parameter. In their notation, $J_c=t_{hop}$, $V=U$, and $\langle N\rangle$ is the number density. The area inside the lobes is a Mott Insulator phase and outside a superfluid phase. The transitions are second order mean field, except at the lobe tips, where they are non mean field. 
  \textbf{Right panel:} The zero-temperature phase diagram of our holographic model with a particular choice of IR potential \protect\eqref{IRP8}, with parameters $\Lambda_{(2,0)}=1$, $\Lambda_{(1,1)}=-3/2$, and other $\Lambda_{(p,q)}=0$. Here $\mu_b = - \mu/V$ and $t_{hop}$ is identified in both diagrams. This IR potential allows for the amplitude of the lobes to decrease inversely with the occupation number. The area inside the lobes is a Mott Insulator phase with the occupation number $\rho$ equal at all lattice sites and a gap of the order of the Coulomb gap, and outside a non homogeneous phase with different occupation numbers at different sites. The non homogeneous phase is also gapped, albeit with a gap much smaller than the Coulomb gap in the system, at least to first order in hopping parameter.  The transition is first order everywhere except at the cusps at $t_{hop}=0$, where it is first order in the $\mu_b$ direction, and second order in the $t_{hop}$ direction. We believe that this difference is an artefact of the hard wall cutoff in our model, and will be remedied once a different background, such as a soft wall like geometry is used. Note that our convention for the sign of $\mu_b$ in \protect\eqref{BHFreeEnergy} differs from the usual sign of $\mu_b$ in \protect\eqref{BHHamiltonian}. 
  }
    \label{fig:FWGFlobe}
  \end{center}
\end{figure}

At fixed nonzero $t_{hop}$, as one increases or decreases $\mu_b$, eventually it becomes energetically favorable to add a particle or hole which can hop around the lattice. At this point, (at zero temperature) the particles will instantly Bose condense and the system will undergo a phase transition to a superfluid. The boundary of the Mott insulating--superfluid phase transition will extend to the $t_{hop}=0$ axis at precisely the points where $\mu_b/U$ is an integer, since at these points the occupation numbers $n$ and $n+1$ are degenerate, and there is no energy cost to adding extra particles. Thus we see that the phase boundaries form a series of lobe shapes extending from the $t_{hop}=0$ axis to some finite value of $t_{hop}$. This can be seen in Figure \ref{fig:FWGFlobe}, taken from \cite{Fisher:1989zza}.

Generically, the transition from the Mott insulating phase to the superfluid phase is continuous and second-order,  driven by the addition of a small number of particles and/or holes to the system, and a change in density from the integer value in the Mott insulating phase to a non-integer value slightly larger or smaller. However, at the tip of the lobes, the Mott insulating phase and the superfluid phase both have integer density $n$. In this case, the transition is driven by an increase in $t_{hop}$ which allows the bosons to overcome the on-site  repulsion and delocalize despite the fact that the density remains constant. As the density $\mu_b/U$ increases, the amount of kinetic energy required to overcome the potential barrier to create one particle/hole pair decreases, and thus the location of the tips of the lobes decrease. This transition is also continuous and second-order, but in a different universality class than the density-driven transition which occurs at all points along the phase boundary except the tips of the lobes.

One can analytically reproduce the phase boundaries of Figure \ref{fig:FWGFlobe} via a mean-field analysis as in  appendix A of \cite{Fisher:1989zza}. Taking an infinite-range hopping limit, they derive an action of the form
\be
S_\infty(\psi)=\beta N\left ( {1\over 2} r(\mu_b, t_{hop},T)|\psi|^2 + u(\mu_b, T)|\psi|^4 + O(|\psi|^6)\right ),
\ee
where $\psi$ is an order parameter which distinguishes between the superfluid, $\langle \psi\rangle =0$, and insulating, $\langle \psi\rangle \neq 0$, phases, $N$ is the total number of lattice sites, and $T$ is temperature. For $T=0$, 
\be
r(\mu_b,t_{hop})= {1\over 2t_{hop}}-{N\over 2} \sum_i \frac{ U(n_i +1)}{Un_i-(t_{hop}+\mu_b)} + \frac{Un_i}{(t_{hop}+\mu_b) - U(n_i-1)},
\ee
and minimization of $S_\infty$ corresponds to the condition $r(\mu_b,t_{hop})=0$. This reproduces the phase diagram of Figure \ref{fig:FWGFlobe}.

\subsection{A holographic model}\label{sec:holographic}

The purpose of this paper is to realize the Bose-Hubbard model via holography. There are many different motivations for studying the Bose-Hubbard model via holography. While exact results are known for the Bose-Hubbard model in low dimensions, the higher-dimensional Bose-Hubbard model is difficult to analyze using field theory techniques alone. It is however possible to study the Bose-Hubbard model using numerical simulations. Such numerical simulations give  quantitative answers, but often little insight into the underlying mechanisms. In contrast, holographic models are at best toy models and so can not be trusted to give quantitatively correct answers for the real Hubbard model, but often give important insights into  non-perturbative mechanisms. Most importantly, implementing the Bose-Hubbard model holographically is a first step towards constructing the dual of the Fermi-Hubbard model. Unlike its bosonic cousin, the Fermi-Hubbard model is not easily accessible with numerical techniques due to the notorious fermion sign problem. It is however of paramount theoretical and practical importance. It is the paradigm for a model of interacting fermions on the lattice and is believed to give a good representation of important materials such as, for example, high T$_c$ superconductors. On the holographic side we do not see the fundamental charge carriers but only the gauge neutral composite operators such as the charge density and $b_i^{a \dagger} b_{ja}$ with $i \neq j$. These composite operators are bosonic irrespective of whether the underlying charge carriers created and annihilated by $b_i^{a \dagger}$ and $b_{ia}$ respectively are bosonic or fermionic. In the bulk, the difference between the Bose-Hubbard model and the Fermi-Hubbard model (or more general models which have both bosons and fermions) should all be in the details of the holographic construction, such as e.g. interactions and boundary conditions. For example, theories with chiral fermions in even dimensions often have anomalous global symmetries, which are reflected in the holographic dual by bulk Chern-Simons terms that are absent in brane constructions with purely bosonic spectra. By focusing first on the Bose-Hubbard model, where we can use the well known phase structure to tune our bulk model, we can establish that holographic techniques indeed work. An obvious next step is to see whether one can generalize our construction to the Fermi-Hubbard model, where holographic techniques have the potential to  add new insights.

%%\begin{table}[t]
%%  \label{TU1}
%%  \begin{center}
%%    \begin{tabular}{|c|c|c|c|c|} \hline
%%Bose-Hubbard  &  Gravity dual  & VEVs in  & VEVs in & Symmetry\\
%%  model (Large $N$)  & &  Sec.~\protect\ref{sec:finitehopping} & App.~\ref{App:MixedNeumann} & Breaking           \\ \hline
%% $\mu$  & $A_{t,i}$  (source) & & & \\ \hline
%%$\rho_i \equiv b_{i}^{a\dagger}b_{i a}$ & $A_{t,i}$  (VEV) & $\neq 0$ & $\neq 0$ & \\ \hline
%%$t_{hop}$  & $\phi_{i,j}$ (source) & & & \\ \hline
%%$b_{i}^{a\dagger}b_{j a}$ & $\phi_{i,j}$(VEV)  & $\neq 0$ & $\neq 0$ & $\text{diag}(U(1)_i \times U(1)_{i+1})$\\ \hline
%%$U$ & hard wall cut-off $r_{h}$ & & & \\ \hline
%%    \end{tabular}
%%  \end{center}
%%  \caption{The AdS/CFT correspondence of the holographic Bose-Hubbard Model. 
%%  The third and fourth column indicate which operators acquire a non vanishing VEV in the non homogeneous phase, for the boundary conditions of Sec.~\protect\ref{sec:finitehopping} and App.~\protect\ref{App:MixedNeumann}, respectively.
%%  }
%%  \end{table}
\begin{table}[t]
  \label{TU1}
  %\begin{center}
  {\small\hspace{-2cm}
    \begin{tabular}{|c|c|c|c|c|} \hline
Bose-Hubbard  &  Gravity dual  & VEVs in  & VEVs in & Symmetry\\
  model (Large $N$)  & &  Mott Phase & Inhomogeneous & Breaking           \\
& & (homogeneous) & Phase &  \\ \hline
 $\mu$  & $A_{t,i}$  (source) & & & \\ \hline
$\rho_i \equiv b_{i}^{a\dagger}b_{i a}$ & $A_{t,i}$  (VEV) & $\neq 0$ & $\neq 0$ & \\ \hline
$\bar \rho \equiv \sum\limits_{i=1}^N {b_{i}^{a\dagger}b_{i a}}$ & $A_{V,t}\equiv \sum\limits_{i=1}^N A_{t,i}$  (VEV) & $\neq 0$ & $\neq 0$ & \\ \hline
$\delta \rho_i \equiv {b_{i+1}^{a\dagger}b_{i+1 a}-b_{i}^{a\dagger}b_{i a}}$ & $A_{A,i,t}\equiv A_{t,i+1}-A_{t,i}$  (VEV) & $= 0$ & $\neq 0$ & \\ \hline
$t_{hop}$  & $\phi_{i+1,i}$ (source) & & & \\ \hline
$b_{i+1}^{a\dagger}b_{i a}$ & $\phi_{i+1,i}$(VEV)  & $= 0$ (Sec.~\protect\ref{sec:finitehopping}) & $= 0$ (Sec.~\protect\ref{sec:finitehopping})& $U(1)_{i+1} \times U(1)_{i} \rightarrow $\\ 
 &   & $\neq 0$ (App.~\protect\ref{App:MixedNeumann}) & $\neq 0$ (App.~\protect\ref{App:MixedNeumann}) & $\text{diag}(U(1)_{i+1} \times U(1)_{i})$\\ \hline
$U$ & hard wall cut-off $r_{h}$ & & & \\ \hline
    \end{tabular}
    }
  %\end{center}
  \caption{The AdS/CFT correspondence of the holographic Bose-Hubbard Model. Here we specialized to nearest-neighbor hopping. The first and second line describe the on-site chemical potential and charge density, the third line the overall charge density, the fourth line the relative charge densities between adjacent sites, the fifth and sixth line the nearest-neighbor hopping, and the last line the Coulomb repulsion parameter. 
  The third and fourth columns indicate which operators acquire a non vanishing VEV in the Mott insulating (homogeneous) and inhomogeneous phases, respectively. Note that whenever there are differences for the boundary conditions of Sec.~\protect\ref{sec:finitehopping} and App.~\protect\ref{App:MixedNeumann}, they are indicated. If the result is the same for both boundary conditions, no distinction is made. Furthermore, note that the VEV (i.e. the normalizable mode in the bifundamental) is chosen to be zero in both phases by the boundary conditions employed in sec. 3, and is nonzero in both phases by the mixed Neumann boundary conditions considered in App.~\protect\ref{App:MixedNeumann}. Nevertheless, the non vanishing source $t_{hop}$ will enact the symmetry breaking indicated in the last column in both phases.}
  \end{table}

In our construction we consider a gauge theory living on an $AdS_2$ background with a hard-wall cutoff. We construct a lattice using a quiver-like gauge theory--we introduce a $U(1)$ Maxwell field corresponding to each site in our lattice, and we couple  bifundamental scalar fields charged under  $U(1)_i\times U(1)_{i+1}$ to the gauge fields at adjacent sites. The hopping parameter will be identified with the source of the operator dual to the bifundamental charged scalar. Its VEV will break the $U(1)_i\times U(1)_{i+1}$ symmetry down to the diagonal subgroup. Charged scalar fields in the bulk can carry the electric (or baryonic) charge and are needed to show the transport of defect (large $N$) bosons between the sites. The hard wall at $r=r_{h}$ is needed to realize the Mott insulating phase with gapped excitations. In the presence of the infrared cutoff (i.e. the hard wall), we can safely consider the effective theory for $AdS_{2}$ in the probe limit~\cite{Maldacena:1998uz,Sen:2011cn}. This holographic  model should be understood to be dual to a Hubbard-like model with $SU(N)$ fundamental bosons localized on each site instead of the spin zero bosons of the original Hubbard model. For this system to have a good holographic dual we are implicitly working in the large-$N$ limit. This large-$N$ limit also helps us theoretically analyze the phase transition  in finite volume.  The dictionary is given in Table~\ref{TU1}, where spin indices are given by $a=1\dots N$.

Our main result is the derivation of a lobe-shaped quantum phase transition structure in the chemical potential - hopping plane, separating a Mott insulating phase from an inhomogeneous phase, similar to the phase diagram shown in Figure \ref{fig:FWGFlobe}. We determine the phase structure by calculating the free energy from the holographically renormalized on-shell action for the phases present, and determine the phase which minimizes the free energy. At zero hopping our model exhibits a level-crossing quantum phase transition which changes the occupation numbers of the bosons between different insulator phases upon variation of the chemical potential.

The order parameter for the insulator--inhomogeneous phase transition is the vacuum expectation value of the operator dual to the bifundamental scalar, which breaks the diagonal subgroup of the $U(1)_i\times U(1)_{i+1}$ global symmetry at neighboring sites. Whether this expectation value vanishes or not depends on the boundary conditions imposed. In sec.~\ref{sec:finitehopping} we impose Dirichlet-like boundary conditions which force the VEV to vanish in the Mott phase but allow it to be nonzero in the inhomogeneous phase. In App.~\ref{App:MixedNeumann} we investigate the Neumann boundary conditions which arise naturally from the variational principle at the hard wall, in which case the VEV will be non vanishing in both phases. As discussed in \S \ref{sec:rev}, in the Bose-Hubbard model this quantum phase transition is of second order. However, we  generically find a first order transition between the superfluid and insulating phases, except at zero hopping (at the cusps between the lobes), where we find a second order transition in the $t_{hop}$ direction. Furthermore, at the cusp points the transitions at zero hopping are first order in the direction of the chemical potential but second order only when we switch on infinitesimal hopping, which is in accordance with the mean-field analysis of \cite{Fisher:1989zza}. One important result is the identification of the ground state of the Mott insulating phase which dominates when $U/t_{hop}$ is large. 

The content of this paper is as follows. In section 2, we consider the case of zero hopping ($t_{hop}=0$). We derive the homogeneous ground state (i.e. same boson density at every site) of the Mott insulator from our holographic model. We furthermore show that the transitions between Mott insulator states with different boson densities are first-order level-crossing phase transitions. In section 3, we consider finite hopping. We holographically derive the Mott insulator/inhomogeneous phase transition in a model with only two lattice sites, showing the lobe-like phase structure. A non-homogeneous state in which the charge densities are not equal on both sites is able to dominate over the homogeneous  state for $t_{hop}\neq 0$.  In section 4, we analyze perturbations around both Mott insulator phase and inhomogeneous phase, to first order in a small hopping expansion. We find two almost zero modes in the inhomogeneous phase and a gap in the Mott insulating phase.  In section 5 we discuss the generalization to the $n$-site model. Finally, in section 6 we discuss a string embedding similar to the Hubbard model and several future directions.

\section{Zero Hopping}

In this section we first analyze our holographic Bose-Hubbard model at zero hopping. We will reproduce the level-crossing transitions reviewed in \S \ref{sec:rev}. The theory is based on a quiver-like lattice of decoupled $U(1)$ gauge theories on a  $AdS_2$ hard wall, with one theory at each site $k$ of the spatial lattice.
The action is simply given by
\ba\label{KIN21}
S_{kin}=\sum_k\int d^2x\sqrt{-g}\Big(-\dfrac{1}{4}F_{(k)\mu\nu}F^{\mu\nu}_{(k)}\Big).
\ea
The metric of the $AdS_2$ hard wall is given by
\ba\label{2e2}
ds^2=-r^2dt^2+\dfrac{dr^2}{r^2},
\ea
where we have set the $AdS$ radius to be 1, and placed the hard wall at $r=r_h$. This means the radial coordinate is restricted to $r\geq r_h$. Note that for the above metric $\sqrt{-g}=1$. This theory is a simplified bottom-up version of the D5 brane lattice of \cite{Kachru:2009xf}, with the internal $S^4$ coordinates as well as the induced brane geometry of e.g. an AdS-Soliton background \cite{Horowitz:1998ha} being neglected. In the following we choose the radial gauge
\ba\label{2e3}
A_{(k)r}=0
\ea
at every lattice site. In the background \eqref{2e2} the Maxwell equations are given by
\ba\label{2e4}
\partial_{\mu}F^{\mu\nu}_{(k)}=0.
\ea
The Maxwell equations are solved by
\ba\label{2e5}
A_{(k)t}=\mu +\rho_{(k)}r
\ea
The same chemical potential $\mu$ is chosen at every site, as will be required by thermal equilibrium when $t_{hop} \neq 0$. In preparation for coupling the lattice sites, we make this homogeneous choice at $t_{hop}=0$ as well, in this way ensuring that the equilibrium configuration at finite hopping is continuously connected to the one at zero hopping. The coefficient $\rho_{(k)}$ is the charge on that site and $\rho_{(k)}=\delta S_{kin}/\delta A_{(k)t}$. As worked out in e.g. \cite{Erdmenger:2013dpa}, only Neumann boundary conditions are possible for a gauge field in $AdS_2$ when coupled to a charged scalar, which fix the charge density $\rho_{(k)}$ and hence force us to work in the canonical ensemble.\footnote{The unit charge is defined by introducing the fundamental charged object as the bulk source term $\int d^2x\sqrt{-g} j^t_{(k)}A_{(k)t}$ with $j^t_{(k)}=\delta (r-r_0)$. The quantization of the charge then depends on the factor of the kinetic term. For our model, $\rho_{(k)}\in \mathbb{Z}$.} For the gauge fields, we impose a Dirichlet boundary condition
\ba\label{2e6}
A_t|_{r=r_h}=u\,,\quad u=\mu+\rho_{{(k)}}r_{h},
\ea
which is the generalized version of the Dirichlet boundary condition $A_t|_{r=r_h}=0$, at the hard wall. It is important to note that since we have no explicit charge carriers in the bulk, the source of charge at infinity is localized at our hard wall. To see this, observe that $F_{rt} \neq 0$ at the wall and so there exists a nonzero electric field normal to the surface at $r_h$. Gauss's law then guarantees the existence of surface charge at the wall by a classical calculation, though the charge is not readily visible in \ref{2e6}. 

Since the on-shell action \eqref{KIN21} has power-law divergences in $r$ at the boundary $r\rightarrow\infty$, we follow \cite{Henningson:1998gx,de Haro:2000xn,Karch:2005ms} and regularize the action by adding the following cutoff term at large $r=R$ (c.f. e.g. the discussion in sec.~3 of  \cite{Erdmenger:2013dpa})
\ba\label{CUT24}
S_{cut}=\sum_k\dfrac{1}{2}\int_{r=R} dt \sqrt{-h}A_{(k)t}A^t_{(k)},
\ea
where $\sqrt{-h}=R$ is the induced metric at the boundary.
A couple of comments on gauge invariance are in order: Note that this counterterm is not manifestly gauge invariant. The boundary term is only invariant under gauge transformations which vanish after the integration over the time direction. As usual in AdS/CFT, we require that suitably quickly decaying gauge transformations do not change the leading coefficient of the boundary expansion of the gauge fields. Only these gauge transformations are truly a redundancy of the theory. Large gauge transformations, that is those that do not vanish at the boundary, change the coupling constants of the theory and so do change the physics. 

What is crucial for our analysis is that the charge on each site is quantized. That is we assume that all charge carriers in the theory have charges which are integer multiples of some basic charge and hence $\rho_{(k)}\in\mathbb{Z}$. This is clearly true in the standard Bose-Hubbard model: the only charge carriers are the bosons and we can take their charge as the basic quantum of charge.
If we assume that our theory can be consistently coupled to monopoles, quantization of both the monopole charge and the electric charge directly follow from the Dirac quantization condition. Any theory that arises as a low energy effective theory for a system made of neutrons, protons and electrons needs to be consistent when coupled to monopoles, as the underlying microscopic theory, QCD and QED, can be consistently coupled to monopoles. Note that this is a purely theoretic statement and true irrespective of whether monopoles actually exist in nature. This argument however only guarantees  quantization of the net charge; what we here assume is the stronger statement that charges are quantized on each site. Again, this is also a condition imposed by fiat in the Hubbard model.

Another, possibly more intuitive argument for charge quantization at each defect can be made from the point of view of probe-brane top-down constructions using the $AdS$ soliton geometry \cite{Horowitz:1998ha}. For more information on such a construction c.f. sec. \ref{sec:Discussion}. Let us e.g. introduce a probe D5-brane on the $AdS_5$  times $S^5$ soliton geometry \cite{Horowitz:1998ha} without considering its back-reaction.  The D5-brane wraps two uncompactified directions and an $S^4\subset S^5$. In addition these branes carry a quantized F1 flux. Such D5-brane embeddings had first been introduced as the dual to the anti-symmetric Wilson loop in $d=4$ $\mathcal{N}=4$ SYM~\cite{Yamaguchi:2006tq,Drukker:2005kx,Hartnoll:2006ib} and are also the main ingredient in the holographic probe brane lattice constructions of refs.~\cite{Kachru:2009xf,Kachru:2010dk}. As shown in \cite{Camino:2001at}, in response to the quantized F1 flux the transverse embedding function of the D5 brane inside the $S^5$ (the azimuthal angle) takes quantized constant values, which leads to a quantization of the energy (tension) the D5 brane can have in different embeddings. The electric components of the worldvolume gauge field flux will then be quantized as well \cite{Camino:2001at}, leading directly to charge quantization on the branes, and to the idea that the D-brane is actually a bound state of a (quantized) number of F1 strings \cite{Camino:2001at}. This ties in nicely with the intuitive picture that the charge on the D-brane corresponds to the number of F1 strings ending on it, which must be quantized. In our bottom-up model we neglect the internal directions crucial for the above argument, and hence we have to impose the charge quantization condition by hand. Note however that the correct limit in top-down models would be the opposite one, in the following sense: The parameter which quantizes the azimuthal angle of the probe D-brane embedding is \cite{Camino:2001at} the ratio $n/N$, where $n$ is the quantized charge. At fixed $n$ in the large $N$ limit the probe D brane hence wraps an infinitesimally small internal sphere, and the DBI description would break down and higher derivative corrections become important. The correct limit to work in for top-down models would hence be the limit of large $n$ and $N$, with their ratio fixed.

Including the counterterms, the holographically renormalized action then becomes $S_{kin}+S_{cut}$. To compute the free energy, we analytically continue~\cite{Hartnoll:2008vx,Hartnoll:2008kx,Donos:2012yu} to Euclidean signature.
 Taking into account the usual minus sign, the free energy is then given by
\ba\label{FEE25}
F=-(S_{kin}+S_{cut})/\beta =\sum_k \Big(\mu \rho_{(k)}+r_h\dfrac{\rho_{(k)}^2}{2}\Big),
\ea
where $\beta=1/T$ is the radius of the time-circle. The plus sign in front of the $\rho_{(k)}^2$ term is important, as it indicates the repulsive Coulomb interactions between bosons at the same site. Conversely, if it were negative, the Coulomb interaction would be negative, and the system could become unstable.

We want to compare the above free energy \eqref{FEE25} with that of the Bose-Hubbard model at zero hopping~\cite{Fisher:1989zza}. Its free energy is given by
\ba\label{BHFreeEnergy}
F_b=\sum_k\Big( \mu_b\rho_{(k)}+\dfrac{1}{2}U\rho_{(k)}(\rho_{(k)}-1)\Big).
\ea
The $\rho_{(k)}^2$ terms describe the on-site repulsive interactions between the bosons.
Matching the parameters $(\mu_b,U)$ with our parameters $(\mu,r_h)$, we find
\ba\label{MAT28}
r_h=U,\quad \mu=\mu_b-\dfrac{U}{2}.
\ea

We conclude that for zero hopping the free energy of our holographic model agrees with that of the Bose-Hubbard model with a repulsive on-site  Coulomb  force.  Moreover, the number operator $\rho_i$ commutes with the Hamiltonian at zero hopping parameter. Thus, the particle number eigenstates are simultaneously diagonalized in the phase with zero hopping, in accord with 
charge quantization at each site $k$. In the Bose-Hubbard model, the homogeneous phase with $\rho_{(k)}=\rho$ (or $A_t^{(k)}=A_t$) at all sites is a Mott insulator for an integer occupation state. We expect the same to be true here, and indeed will show this below in sec.~\ref{sec:levelcross} by directly evaluating the free energy of the different phases, and also  in section~\ref{sec:4} by analyzing the fluctuations around this state. We will in particular find that the gap is given by the Coulomb parameter $U$. Hence our state is a Mott insulator where the fundamental bosons localize because of the on-site Coulomb energy $U$, set by the gap scale $r_h$ of the AdS soliton \cite{Horowitz:1998ha}.
 It is a non-compressible state with gapped excitations and vanishing quasi-particle density of states at energies below the gap scale. In fact, we find that the whole excitation spectrum in the Mott phase is discrete and gapped, as expected from a system in a AdS hard wall geometry.

\subsection{Level-crossing First Order Phase Transitions}\label{sec:levelcross}

\begin{figure}[htbp]
  \begin{center}
   \includegraphics[height=7cm]{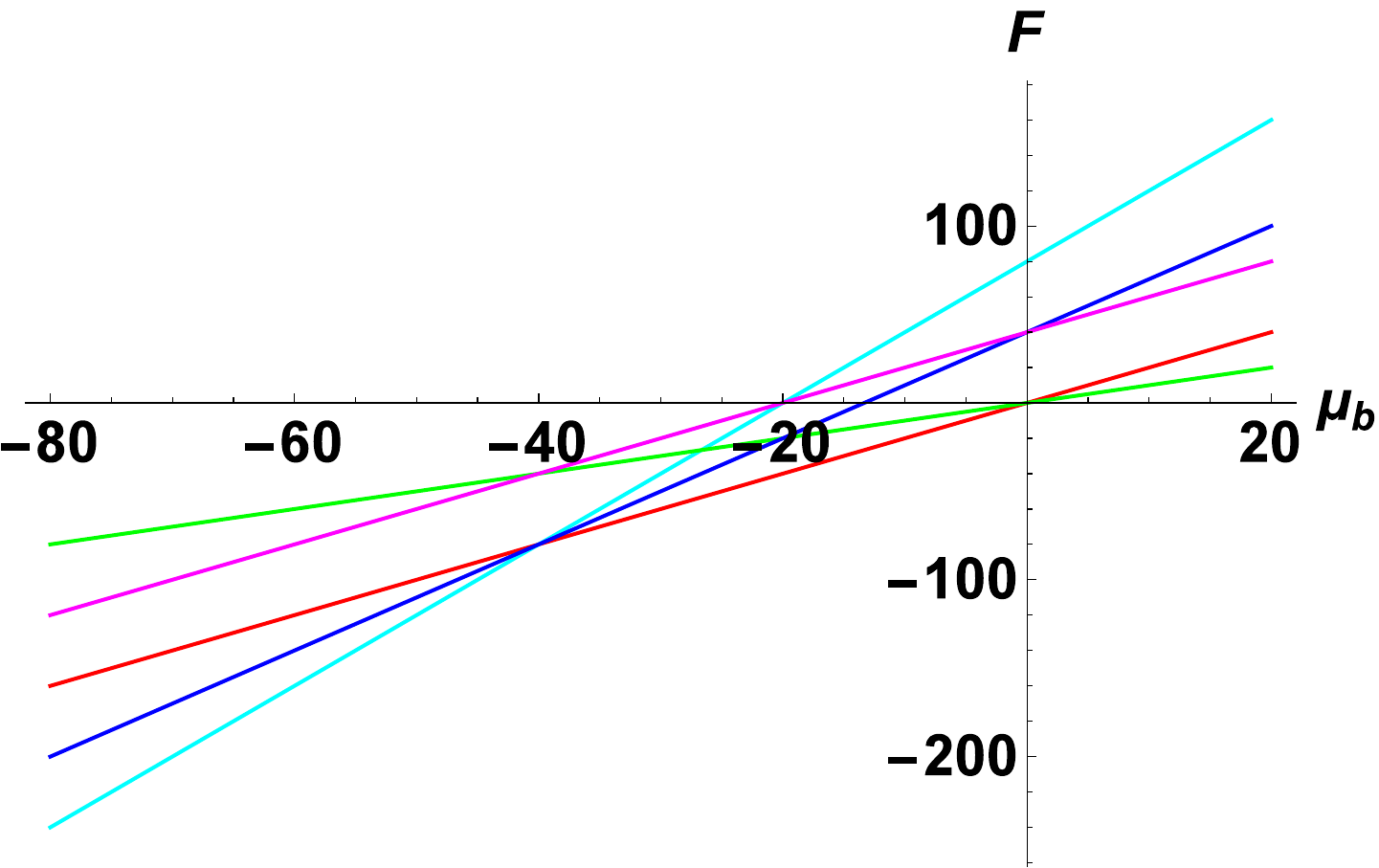}
  \caption{The free energy of the two-site model at zero hopping as the function of the chemical potential $\mu$. We plot the following lines by color:  $(\rho_{(1)},\rho_{(2)})=(0,0)$ (black), $(0,1)$ (light green), $(1,1)$ (red), $(2,0)$ (purple), $(1,2)$ (dark blue), and $(2,2)$ (light blue). The free energy for $(\rho_{(1)},\rho_{(2)})=(0,0)$ is zero, so the black line is flush with the $F=0$ axis. Note that the free energy of the non-homogeneous phase ($\rho_{(1)}\neq \rho_{(2)}$) is symmetric under $\rho_{(1)}\leftrightarrow \rho_{(2)}$, i.e.  these states are degenerate.}
    \label{fig:Level}
  \end{center}
\end{figure}

In the following we analyze the phase structure of our system at zero hopping. We focus in particular on the two-site model, and comment shortly on the (at zero hopping completely analogous) multi-site case at the end of this section. Note that in order for more fundamental bosons to occupy each site we should, according to \eqref{FEE25}, consider negative chemical potential. In this case of negative chemical potential, we can compare our analysis with the phase structure of the Bose-Hubbard model at strongly interacting region $U/t_{hop}\gg 1$.

In Fig. \ref{fig:Level}, we plot the free energy as the function of $\mu_b/U$, where for convenience we chose $U=r_h=40$ in all our numerical calculations in the rest of this paper. Although this choice seems to be arbitrary at first sight due to the rescaling invariance of of $AdS_2$, it is important not to choose the cutoff too small in order to avoid the appearance of other possible instabilities at energy scales above $U$. Since the occupation number is fixed to be an integer, there is a first order transition called the level-crossing phase transition between two Mott insulator phases at the critical values
\be\label{criticalmu}
\mu/U + \frac{1}{2} \equiv \mu_b/U=0,-1,-2,\dots\,.
\ee
 Decreasing $\mu$ induces a level-crossing transition where the occupation number at each  site increases by 1.\footnote{This first order phase transition is similar to that of~\cite{Kobayashi:2006sb,Mateos:2007vc} where the phase transition is from zero density to finite density as the chemical potential increases.} Note that, in the figure, the free energies of the systems that are unequally occupied such as $(\rho_{(1)},\rho_{(2)})=(0,1)$ and $(1,2)$ have degenerate energies at the critical chemical potential \eqref{criticalmu}. At the phase transition point, the free energy at each site is the same for two different occupation numbers and as the sites are all independent each can individually chose between the two options. Unequally occupied states are never preferred at zero hopping; they always have a larger free energy than the preferred state away from the phase transition points.

As seen in the parameter matching, this phase structure is the same as the phase structure of the Bose-Hubbard model in the strongly interacting region. We see that the ground state is described by a Mott insulator.
Without hopping, we can easily generalize the level-crossing phase transition to the case of many sites.
We find a phase structure similar to the two-site model. At the critical chemical potential, the number of bosons in the degenerate state is different by 1 and the whole system has a $2^M$ degeneracy where $M$ is the number of the lattice site. The large degeneracy can be interpreted as a macroscopic entropy.

\section{Finite Hopping}\label{sec:finitehopping}
In the holographic theory we add bi-fundamental scalars to describe the hopping parameter on the field theory side.
As indicated in Table~\ref{TU1}, this bi-fundamental scalar is dual to the operator
$b_{i}^{\dagger}b_{j}$ in the Bose-Hubbard model, which is exactly the term we add to the Hamiltonian when including hopping.
We then analyze the motion of fundamental bosons of the Bose-Hubbard model between sites as a function of the hopping parameter. We mostly consider a simple two-site model ($k=1,2$) and only briefly comment, in the end, on the generalization to a multi-site theory. The bi-fundamental scalar $\phi$ is charged under $U(1)^2$ as $(q,-q)$. \footnote{Note that this bi-fundamental scalar is neutral under the diagonal $U(1)$ subgroup, which is the $U(1)$ that couples to net electric (baryonic) charge. If we want to describe, in the bulk, a charged order parameter we need to, in addition, introduce fundamental scalars charged as $(0,\pm q)$ or $(\pm q,0)$.}
 The action of the bi-fundamental scalar is given by
\be\label{IRP312}
S_{matter}=-\int d^2x\sqrt{-g}|D \phi|^2-\int_{r=r_h} dtr_h {\Lambda} (|\phi|^2+w^2)^2,
\ee
where $D_{\mu}=\partial_{\mu}-iqA_{\mu}^{(1)}+iqA_{\mu}^{(2)}$. The last term is the IR potential describing the interaction of the bi-fundamentals. $w$ is a constant which gives an IR mass for the fields; such a term has been analyzed in~\cite{Csaki:2003dt,Csaki:2003zu}. In our case, a proper choice of IR potential will be necessary to reproduce the quantum phase transition structure of the Bose-Hubbard model at small hopping.

Since we are interested in a static configuration, we consider fields depending only on the $AdS$ radial direction $r$, namely, $A_{t}^{(i)}(r),\ \phi (r)$.
The equations of motion (EOMs) following from the total action are then given by
\ba\label{EOMinhom}
&(r^2\phi')'+\dfrac{q^2}{r^2}(A_t^{(1)}-A_t^{(2)})^2\phi=0,\nonumber \\
&(A_t^{(l)\prime})'-\dfrac{2q^2|\phi|^2}{r^2}(A_t^{(l)}-A_t^{(l+1)})=0,  \label{EOE39}
\ea
where $l=1,2$. Henceforth we choose the charge $q$ to be \be\label{qchoice} q=\sqrt{6}/5\,.\ee
There are several reasons for this choice: First, we should choose $q^2$ large enough to ensure as few subleading corrections to the $t_{hop}$ term in \eqref{BIF314} as possible before the VEV term sets in, making the extraction of the VEV numerically simpler.  Secondly, for the lowest possible value $\delta\rho=1$ in \eqref{BIF314}, the particular choice of \eqref{qchoice} leads to a rational dimension of the operator dual to $\phi$, \be\label{phidim}\Delta_\phi = \frac{3}{5}\,.\ee Finally, a large $q^2$ also ensures that the probe limit for the gauge field and the bifundamental is valid even for small values of $r_h$. When the charge $q$ is much larger than the gravitational coupling constant, we can ignore the back-reaction onto the metric~\cite{Herzog:2009xv,Horowitz:2008bn}. We assume this to be the case and completely neglect the gravitational sector of the bulk theory in this work and treat the background metric as fixed. Note that the diagonal gauge field $A_V=A^{(1)}+A^{(2)}$ decouples from the axial sector $A_A=A_t^{(1)}-A_t^{(2)}$ and $\phi$.

\subsection{Homogeneous Mott Insulator}

When the homogeneous phase $A_t^{(1)}=A_t^{(2)}$ is considered, the EOMs of the fields $\phi$ and $\ A_t^{(l)}$ become independent. We can then solve the EOMa of the fields \eqref{EOE39} analytically. The solutions become
\be\label{SOP310}
\phi=t_{hop}+\dfrac{\varphi_0}{r},\quad A_t^{(l)}=\mu+\rho_{(l)}r,
\ee
where $l=1,2$. We identify the coefficient $t_{hop}$ and the coefficient of the normalizable mode $\varphi_0$ with the hopping parameter and vacuum expectation value (VEV) of the bi-local field $b_{i}^{\dagger}b_{j}$ in Bose-Hubbard model, respectively.

We choose the Dirichlet boundary condition \be\phi|_{r=r_h}=K\,,\label{Dirichletphi}\ee similarly to the gauge field. We set to zero the VEV of the bi-fundamental by tuning the parameter $K$ such that \be \phi|_{r=r_h}=K=t_{hop}\,,\label{Kchoice}\ee while the hopping parameter $t_{hop}$ is used as a free parameter.

In the homogeneous phase, we do not need to add counter-terms for the action $S_{matter}$. The free energy becomes
\be\label{FMottIRPot}
F_{Mott}=-(S_{kin}+S_{cut}+S_{matter})/\beta =2\mu\rho_{(1)}+r_h\rho_{(1)}^2 +r_h\Lambda (t_{hop}^2+w^2)^2.
\ee
Note that at zero hopping the last term is simply a constant shift of the free energy, and hence does not affect the identification \eqref{MAT28}. At
finite hopping this additional term becomes crucial for the physics of our model, in particular, for the existence of the cusps in the phase diagram fig.~\ref{fig:Lobe}.

\subsection{Non-homogeneous Mixed State}

When we consider the non-homogeneous case $A_t^{(1)}\neq A_t^{(2)}$,
an analytic solution to the equations of motion \eqref{EOMinhom} does not exist in general, so we solve \eqref{EOMinhom} numerically. {Although the axial $U(1)$ is explicitly broken by the hopping parameter, and hence the system does not admit an actual Goldstone mode, the non-homogeneous phase still is quite similar to the superfluid phase of the Bose-Hubbard model in the following sense: As we will show in sec.~\ref{sec:nonhompert}, the phase of the VEV of the kinetic energy operator dual to the bifundamental $\phi$ can be related to the superfluid quantum current flowing between different lattice sites in the Bose-Hubbard model in a straightforward way. As it turns out, the phase of the VEV of the kinetic energy operator is completely determined by  the normalizable mode $\varphi_v$ in the bifundamental bulk field $\phi$ in \eqref{BIF314}, as the other contributions to the variation of the free energy with respect to the hopping parameter are manifestly real. Although in the body of this work we are going to choose a boundary condition such that this quantum current (and hence the normalizable mode $\varphi_v$) vanishes, it is naturally present for more generic boundary conditions such as the ones discussed in App.~\ref{App:MixedNeumann}. We hence would like to think of the non-homogeneous phase as a superfluid-like phase, similar to the actual superfluid phase in the Bose-Hubbard model. Note also that  the finite hopping parameter forces the system to live in a state with unequal charge densities, hence possibly forcing additional bifundamental fields to condense spontaneously. If order parameters transforming as fundamentals under one of the gauge groups are coupled to this system in the right way, similar symmetry breaking patterns may arise as well. We are going to comment on both possibilities further below. Although below we are going to work with a charge non-homogeneity of $\delta \rho=1$, at larger $t_{hop}$, higher and higher $\delta \rho$ will presumably be the dominating phase. Also, if we had chosen boundary conditions which allow for a bifundamental VEV, such as the ones in App.~\ref{App:MixedNeumann}, the VEV would grow asymptotically large with larger $t_{hop}$.}  

The solutions of the EOMs \eqref{EOMinhom} satisfy the following UV asymptotics:
\ba\label{BIF314}
&\phi \sim t_{hop}r^{\alpha_t}-\dfrac{4  \delta\rho^2 q^4
     t_{hop}^3r^{
    {3} \alpha_t}}{(2\alpha_t+1) \alpha_t (q^2\delta\rho^2 +
      {3} \alpha_t +
      {9} \alpha_t^2)} + {\cal O}(r^{5\alpha_t}) + \varphi_{v} r^{-1-\alpha_t}(1 + \dots) , \nonumber \\
      &A_t^{(l)}\sim\mu +\rho_{(l)}r-(-1)^l\dfrac{\delta\rho q^2 r^{2\alpha_t+1} t_{hop}^2}{
 (2\alpha_t+1) \alpha_t} + {\cal O}(r^{4 \alpha_t + 1}),
\ea
where $\delta \rho=\rho_{(1)}-\rho_{(2)}$ and $\alpha_t=(-1+\sqrt{1-4q^2\delta\rho^2})/2$. In the above asymptotic expansion, the subleading corrections due to the hopping parameter are included since this term becomes important in both the asymptotic expansion of the gauge fields and of the bifundamental. For example, there are finitely many correction terms to the hopping term in the expansion of $\phi$ before the normalizable mode $\sim \varphi_v$ takes over. The same is true in the gauge field expansion.\footnote{Note that in this paper we use the notation $\varphi_v$ for the normalizable piece in the inhomogeneous phase,  $\varphi_0$ for the normalizable piece in the Mott phase, and $\tilde \varphi = (1-2\Delta_\phi)\varphi_v$ for the actual value of the VEV associated with the normalizable mode in the inhomogeneous phase (c.f.~App.~\ref{App:MixedNeumann}).}  To stay stable, i.e. above the BF bound
\be\label{BF}
4 q^2 \delta \rho^2 \leq 1\,,
\ee
 that is to keep $\alpha_t$ real, we can only consider the case $|\delta \rho |=1$ for the integer occupations $\rho_{(l)}$ and our choice of charge \eqref{qchoice}. Note however that the restriction \eqref{BF}, which does not exist in the usual Bose-Hubbard model, is not a mere technicality, but arises from the coupling of the Bose-Hubbard model to a large N CFT, which is implicit in our holographic setup.\footnote{We thank the anonymous referee for this comment.} This  introduces an extra parameter, the 't Hooft coupling $\lambda$, as an additional direction in parameter space. At weak 't Hooft coupling, the theory would even make sense if \eqref{BF} was violated, and such instances have for example been studied in \cite{Pomoni:2008de, Pomoni:2010et, Kaplan:2009kr, Kutasov:2011fr}. But then the same theory will be unstable at strong coupling with a hitherto unknown actual ground state, and hence we choose our parameters so that the dimension of the scalar sits above the unitarity bound, and the theory is stable. The bi-fundamental scalar was introduced as the holographic dual to the bi-local field $b^{\dagger}_{i}b_{j}$. In particular, the identifications of $t_{hop}$ and $\varphi_{v}$ as the hopping term and the dynamically generated part of the vev of this operator should still hold in the non-homogenous phase. We can interpret $\alpha_t$ as quantum corrections to the dimensions of this operator (i.e. its anomalous dimensions) from the interactions in the non-homogeneous phase.

\begin{figure}[htbp]
  \begin{center}
   \includegraphics[height=7cm]{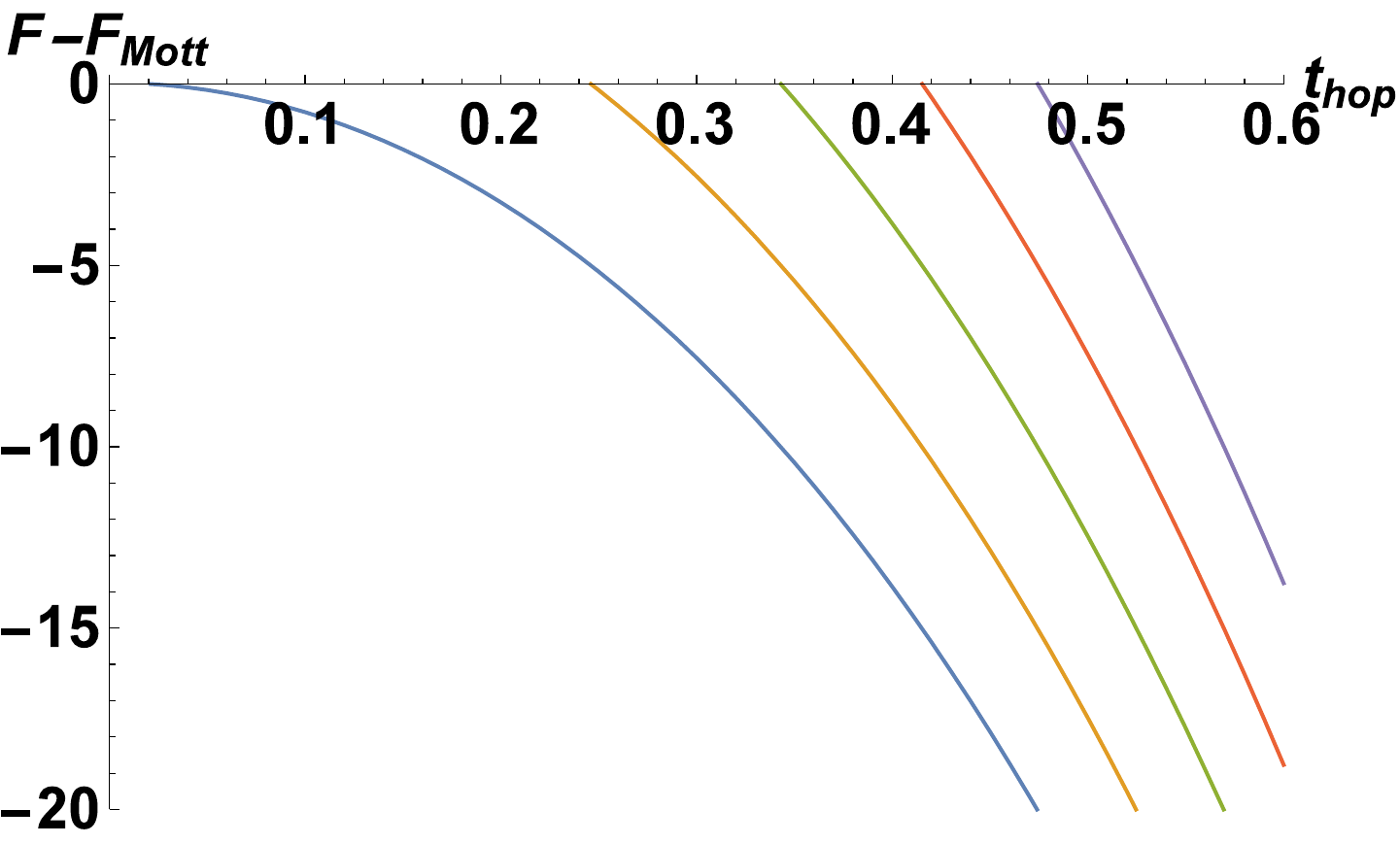}
  \caption{
  The difference of the free energy $F-F_{Mott}$ between the non-homogeneous phase and the Mott insulator phase $(\rho=2)$ as the function of $t_{hop}$. From the left to the right, the line represents the free energy with $\mu_b/U=\mu/U +1/2=-1,-1.125,-1.25,-1.375,-1.5$, respectively. The free energy with $\mu_b/U=-1$ shows that the non-homogeneous phase always dominates when $t_{hop}\neq 0$. After crossing the particle-hole symmetric point    $\mu_b/U=-1.5$, the curves are attained again in reversed order, until the next cusp point at $\mu_b/U=-2$ is reached. Note that in our model at the particle-hole symmetric points $\mu_b/U = -(n+1/2)$, the three phases $(n,n)$, $(n,n-1)$ and $(n+1,n)$, compete, and  the transition is first order even in the non-homogeneous phase (c.f. fig.~4). 
  }
    \label{fig:Lobew1free}
  \end{center}
\end{figure}

We choose the boundary condition in which the subleading term $\varphi_{v}$ in the $AdS$ boundary expansion vanishes. For real $t_{hop}$, as we choose throughout, the imaginary part of $\varphi_{v}$ is related to the quantum current of our theory, as discussed in section 4. Our choice of boundary conditions hence ensures a vanishing quantum current in the ground state. The requirement that the real part of $\varphi_{v}$ also vanishes is however a stronger constraint. The reason for choosing this boundary condition can be motivated as follows. Our system does not exhibit purely spontaneous breaking of $U(1)_1-U(1)_2$. The symmetry is always explicitly broken by the presence of the source, $t_{hop}$, and the VEV generated by it (c.f.~fig.~\ref{fig:Lobew1free} and eq.~\eqref{FRE315}). This VEV turns out to be non-zero even when $\varphi_v$ vanishes, as the VEV is given by the variation of the on-shell action with respect to the source $t_{hop}$, which contains contributions both from $\varphi_{v}$ \textit{and} from counterterms. The latter will not vanish even when the former does. We take the ground state to be the one with vanishing current, so we can safely set the imaginary part of $\varphi_v$ to zero. Further, we want to interpret setting all of $\varphi_v$ to zero as demanding that the $(U(1)_1 - U(1)_2)$-breaking VEV is entirely forced upon us by the source and has no spontaneous component, which would correspond to the real part of $\varphi_{v}$. Of course, there is no rigorous way to break up the VEV this way and so this argument can at best serve as a heuristic motivation for our choice. 

Note that in this way we specify two UV boundary conditions in solving the model, and no boundary conditions at all at the IR hard wall. This is  different from how one usually proceeds in AdS/CFT. We checked that our fields behave regularly at the hard wall, so in principle we can always reinterpret this UV boundary condition as a particular IR boundary condition. Whatever value the field takes on the IR wall could be viewed as an input that the ensures vanishing of $\varphi_{v}$. In the case of IR Neumann boundary conditions, which we analyze in App.~\ref{App:MixedNeumann}, one can view vanishing of $\varphi_{v}$ as a particular choice of potential on the hard wall.\footnote{Since the hard wall itself does not fulfill the background Einstein equations, it has a certain energy and momentum induced  from this nonfulfillment of the EOMs, and since we don't know this additional source of energy-momentum, we should not worry too much about the IR boundary contributions coming from the hard wall in the first place. The boundary conditions will be determined in principle in a top-down construction where the background explicitly solves the supergravity equations of motion, and most probably induces an $AdS_{2}$ hard wall-like geometry on the brane.}

It should probably be noted at this point that by imposing charge quantization with order one charges on the gauge field, we essentially chose two UV boundary conditions in this case as well. We specify the chemical potential (the source) but then only allow the charge to  take a few discrete values. Both of those are UV data. In a top-down AdS/CFT setup classical equations in the bulk are only valid when the quantized charge is large (or order $N$). In this case the charge density can be treated as a continuous parameter in the UV, which needs to be fixed (as usual) by an IR boundary condition. By insisting on order one charges for the gauge field we break the standard rules of AdS/CFT. This could potentially be justified by working at large but finite $N$. Since we are forced into this situation for the gauge field already, we should probably not be surprised that we need to follow a similar strategy for the scalar as well. Without imposing vanishing of $\varphi_v$ we were unable to produce the nice lobe structure we present in here.

\begin{figure}[htbp]
  \begin{center}
   \includegraphics[height=7cm]{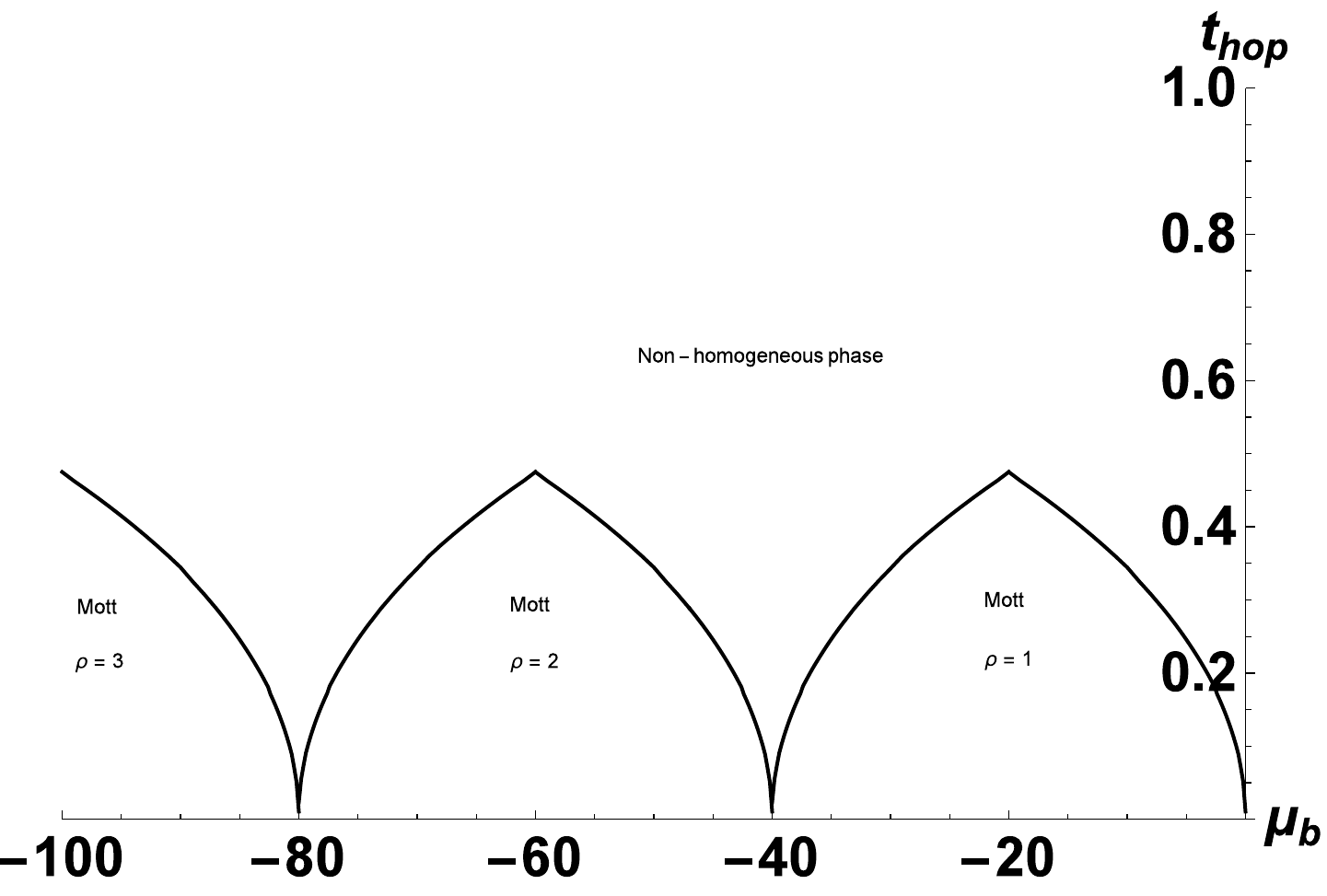}
\caption{
The phase structure of the two-site model in the ($\mu_b,t_{hop}$)-plane for $r_h=40$, $w^2=1$, and $\Lambda =1$. Note that we use the chemical potential $\mu_b=\mu+ U/2$. Inside the lobes, the charge density on both sites is equal, and by analyzing the spectrum of excitations (c.f.~sec.~4) we identify this homogeneous phase with the Mott insulating phase. As $t_{hop}$ is increased, there are regions (which we call inhomogeneous phases) where the non-homogeneous states are favored. Note that the width of the lobe is fixed in units of $U$, basically by the free energy at zero hopping. According to our experience, changing the $w$ parameter (an IR mass) in the IR potential, the height of the peaks behaves as $t_{hop}\sim 1/w$. The cusps appear due to the degeneracy of the ground state between homogeneous phase and non-homogeneous phase at the special points on the $\mu$-axis.
}
    \label{fig:Lobe}
  \end{center}
\end{figure}

\begin{figure}[htbp]
  \begin{center}
   \includegraphics[height=7cm]{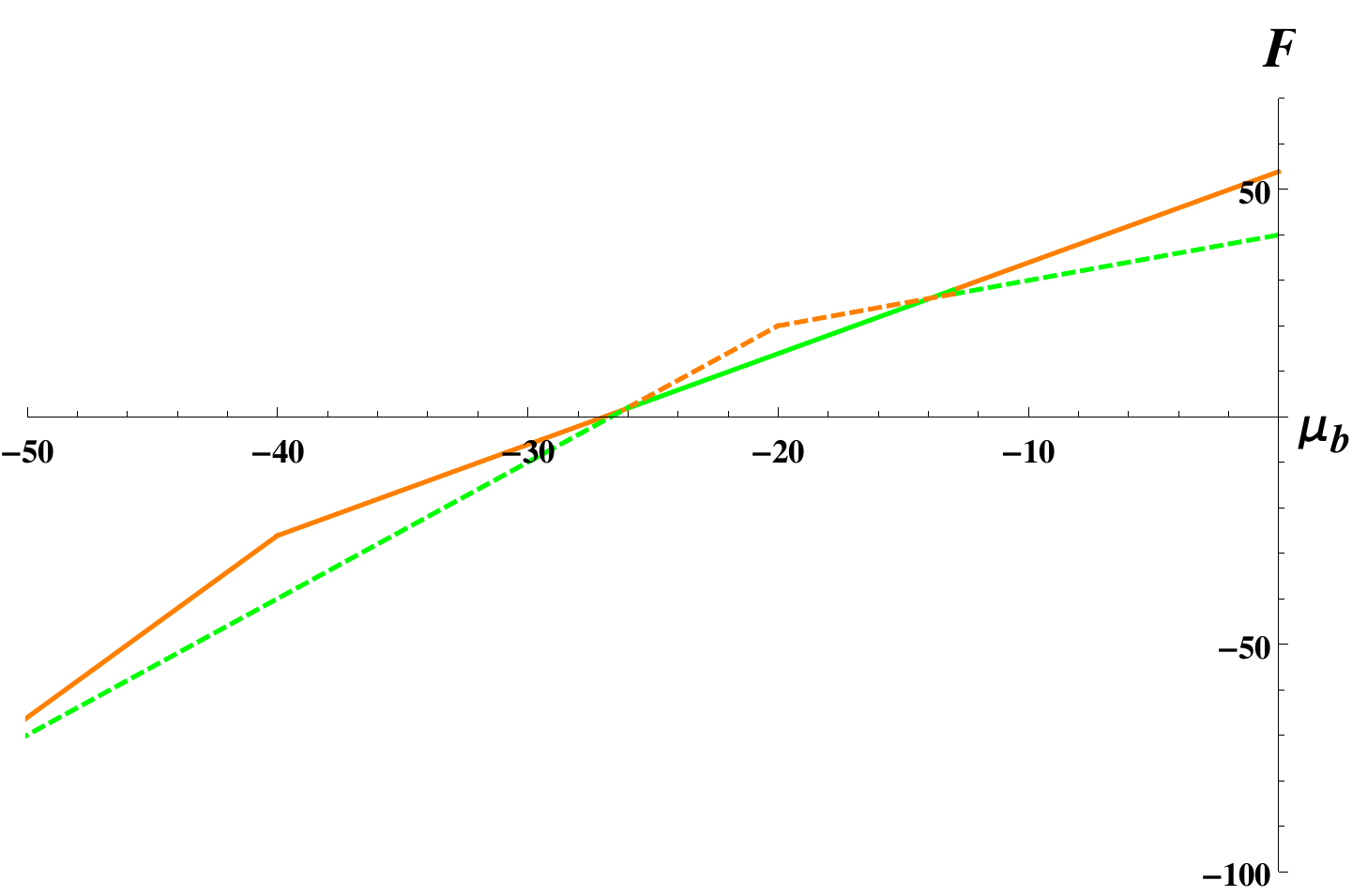}
  \caption{
  The free energy is plotted as the function of $\mu_{b}$ for $t_{hop}=0.4$.  The color coding is as follows: Green lines correspond to the thermodynamically favored phase, while red lines to unstable phases of higher free energy. Solid lines correspond to the Mott phase, dashed lines to the inhomogeneous phase. The  Mott insulator phase is hence the stable ground state for most values of $\mu_b$ at this value of the hopping parameter. The Mott insulator is unstable in some regions of the chemical potential, between the lobes, where the inhomogeneous phase takes over. The free energy hence reflects the lobe structure in Fig. 3.  
}
    \label{fig:Lobeche}
  \end{center}
\end{figure}

\noindent
Since the on-shell action is divergent for the asymptotics \eqref{BIF314}, we add the following counter-terms
\ba
S_{cut,2}=-\alpha_t\int_{r=R} dt\sqrt{-h}\phi^2.
\ea
Then, the free energy is evaluated by the following sum:
\ba
F=-(S_{kin}+S_{matter}+S_{cut}+S_{cut,2})/\beta .
\ea
Because the diagonal gauge field $A_V=A^{(1)}+A^{(2)}$ decouples from the remaining fields, we can rewrite $F$ as
\be\label{FRE315}
F=\mu \sum_i\rho_{(i)}+E\Big(\sum_i\rho_{(i)},\delta\rho, t_{hop}\Big).
\ee
This implies that the energy, which is the second term in the above equation, does not depend on the chemical potential at zero temperature. 
The first few orders of the strong coupling expansion in terms of $t_{hop}$ and $1/U$ for the energy is given by
\ba E(\rho_{(1)},\rho_{(2)},t_{hop})=U\sum_{i}\dfrac{\rho_{(i)}^{2}}{2}+U^{\frac{1}{5}}\Big(-\dfrac{13}{5}+2\Lambda w^{2} \Big)t_{hop}^{2}+\dots \,.\ea
 where we used $|\delta\rho |=1$ and dots include a constant shift and higher order terms. The second term is the leading order $\phi^{2}$ contribution to the energy (see also ~\cite{Bao:2013fda}). 

In Fig. \ref{fig:Lobe} we plot the phase structure of the two-site model for $r_h=40$, $w^2=1$, and $\Lambda =1$. The chemical potential  $\mu_b$ is defined in \eqref{MAT28}. We see that our model reproduces the lobe-like structure of the Bose-Hubbard model. We have regions where the inhomogeneous state is favored at finite $t_{hop}$. In Fig. \ref{fig:Lobe}, the inhomogeneous state extends to the $\mu$-axis at $\mu_b/U\equiv\mu/U +1/2=0,-1,-2$. The width of the lobe is  fixed in units of $U$. Note that the cusps appear wherever the homogeneous and inhomogeneous states have degenerate free energy.

Our experience shows that (at least for the parameter ranges we explored) we can change the amplitudes of all lobes by changing this parameter $w$. In particular, we find that large $w$ decreases the amplitude as $t_{hop}\sim 1/w$.
In Fig. \ref{fig:Lobe}, the amplitudes of the lobes do not change as we decrease $\mu$, while in the actual Bose-Hubbard model  ~\cite{Fisher:1989zza}, the amplitudes decrease as $1/\rho_{(1)}$, due to the fact that as the number density of the sites increases, one needs less kinetic energy to overcome the potential barrier of removing a particle from one site. We will expand on how to reproduce this feature of the lobe structure in the next subsection, sec.~\ref{sec:decrease}.

In Fig. \ref{fig:Lobeche}, we plot the free energy as a function of $\mu_b$ for the fixed hopping parameter $t_{hop}=0.4$. There, the green lines are the thermodynamically prefered phases, while the red lines are the nonprefered phases with higher free energy, and the solid line corresponds to the Mott phase, while the dashed line corresponds to the inhomogeneous phase. Note that fig.~\ref{fig:Lobeche} is completely consistent with the lobe structure in Fig. \ref{fig:Lobe}, that is the values of $\mu_b$ where the lines of free energy cross precisely correspond to the boundaries of the lobes in Fig \ref{fig:Lobe}.

In Fig. \ref{fig:Lobew1free} 
 we plot $F-F_{Mott}$  
as a function of $t_{hop}$. As seen in Fig. \ref{fig:Lobew1free}, the Mott insulating phase is dominating the thermal ensemble in the small $t_{hop}$ regime, while the non-homogeneous phase is dominating in the large $t_{hop}$ regime, as expected.
Finally, in Fig. \ref{fig:Lobew1free}, we see the existence of a second order phase transition near $t_{hop}=0$, reproducing the behavior of the Mott/Superfluid transition in the Bose-Hubbard model at the cusps at small hopping.\footnote{Note however that in order to decide whether our model admits a Goldstone mode, the typical sign of superfluidity, further numerical analysis, in particular of the boundary conditions employed in App.~\protect\ref{App:MixedNeumann} will be necessary. For more comments on this, c.f. sec.~\ref{sec:Discussion}.} However, our model shows a first order phase transition between the Mott phase and the inhomogeneous phase for any $t_{hop}$ except $t_{hop}=0$. First order transitions are more common in holographic large $N$ theories, where they often arise from the free energy competition of several saddle points, so the appearance of a first order transition in this model should not be surprising. For comments on how to achieve a continuous phase transition in this model, c.f.~Sec.~\ref{sec:Discussion}. 

\begin{figure}[htbp]
  \begin{center}
   \includegraphics[height=7cm]{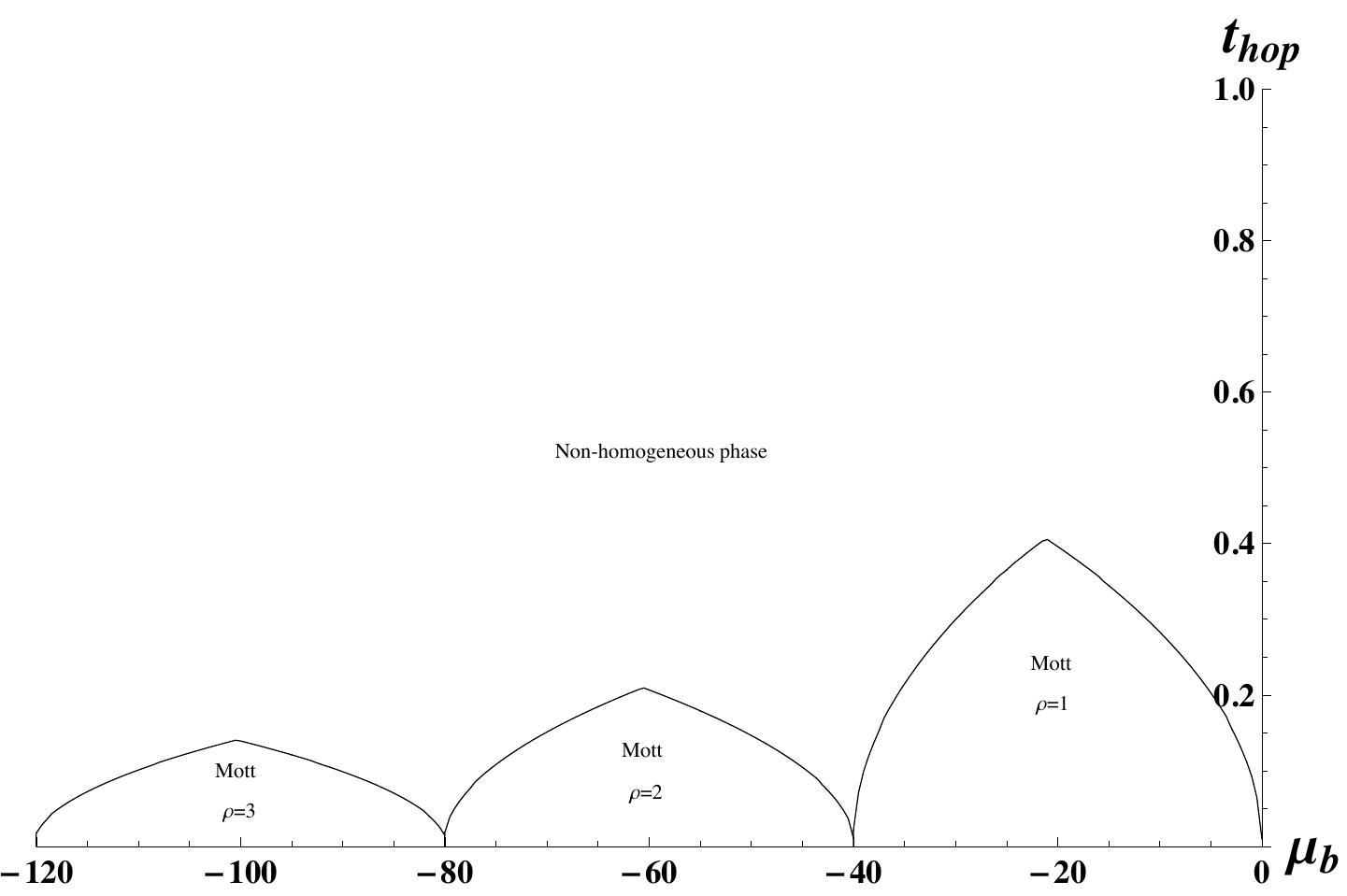}
  \caption{
  The lobe-shape structure of the holographic model with the generalized IR potential. The parameters are chosen to be $\Lambda_{(2,0)}=1$, $\Lambda_{(1,1)}=-3/2$, and other $\Lambda_{(p,q)}=0$. The phase structure shows the decreasing amplitude of the lobe like $1/\rho_{(1)}$. Compare with Fig. 1.
  }
    \label{fig:Lobedec}
  \end{center}
\end{figure}

\subsection{Decreasing the Amplitude of the Lobe}\label{sec:decrease}

In the previous sections, we used the IR potential~\eqref{IRP312} without a gauge potential. To realize the decreasing amplitudes of \cite{Fisher:1989zza} and shown in Fig. \ref{fig:FWGFlobe}, we need to deform our holographic model. In this section, by coupling the gauge fields with the IR potential~\eqref{IRP312}, we reproduce the $1/\rho_{(1)}$ behavior of the height of the lobe \cite{Fisher:1989zza} via holography.

 The IR potential can be generalized to the following gauge-invariant IR potential: \footnote{We choose the IR potential without the gauge potential $A_{t}$ since it breaks the gauge invariance in general.}
\ba\label{IRP8}
&S_{IR}(\phi,F_{rt})=-\int_{r=r_{h}}dt r_{h}\Big(\Lambda_{(1,0)}\abs{\phi}^{2}+\Lambda_{(2,0)} \abs{\phi}^{4}+\sum_{i}\Lambda_{(1,1)}\abs{\phi}^{2}{F_{\mu}^{(i)}F^{(i)\mu}}\nonumber \\
&+\dots +\Lambda_{(p,q)}\abs{\phi}^{2p}\sum_{i}(F_{\mu}^{(i)}F^{(i)\mu})^{q} \Big),
\ea
where $F_{\mu}^{(i)}=F_{\nu\mu}^{(i)}n^{\nu}$ ($n^{\mu}$ is the boundary normal satisfying $n_{\mu}n^{\mu}=1$) and $F_{\mu}^{(i)}F^{(i)\mu}=-F_{rt}^{(i)2}$. To connect with the level changing transitions in the $t_{hop}=0$ case, we require the IR potential to be a purely additive constant to the free energy at zero hopping. In particular it should not explicitly depend on the chemical potential $\mu_b$. It can however depend on the charge density in the diagonal sector, i.e. through the quantized field strength $F_{V}=\sum_{i}\rho^{(i)}$, which does not depend on the radial direction. We plot the phase structure of the model with the above potential in Fig.~\ref{fig:Lobedec} by setting $\Lambda_{(2,0)}=1$, $\Lambda_{(1,1)}=-3/2$, and other $\Lambda_{(p,q)}=0$.
Note that \ref{IRP8} with this choice of parameters is similar to \ref{IRP312}, except that $(\sum_i\rho_i^2)$ now plays the role of the bifundamental mass $w^2$, which controls the location of the lobes' apexes. The height of the apexes behaves as $t_{hop}\sim 1/w$, as already stated in the previous section.
 The difference is that the $w^4$ term is missing. Hence, the difference in free energies between the homogeneous and non-homogeneous phases varies with the on-site occupation number like $\sim 1/\rho_{(l)}$, as desired. Note also that we chose $\Lambda_{(1,1)}=-3/2$ to help achieve this specific functional dependence on $\rho$.
Since the IR potential does not affect the EOM, the perturbations and the excited spectrum in section \ref{sec:4} are not affected by this change due to our choice of free boundary conditions in the IR. In App.~\ref{App:MixedNeumann} we  analyze the mixed Neumann boundary conditions that follow from the variation of the action with the choice of IR potential \eqref{IRP312}, and show that the structure of the phase diagram of our model is qualitatively unaffected by this change.

\section{Perturbations and Excitation Spectrum at Small Hopping}\label{sec:4}

In this section, for small hopping parameter, we compute the perturbation around the background of both the Mott insulator phase and the non-homogeneous phase. We show that for Dirichlet boundary conditions at the IR wall, we find no zero modes in the Mott insulator phase, consistent with the existence of a gap.  We show that two almost zero modes ($\omega\ll r_{h}$) appear in the non-homogeneous phase.% though we do not find a Goldstone mode consistent with superfluidity.% or the usual sign of spontaneous symmetry breaking, the Gell-Mann-Oakes-Renner relation \cite{GMOR}.

The EOMs of the total bottom-up action $S_{kin}+S_{matter}$ \eqref{KIN21} and \eqref{IRP312} are given by
\ba
&\partial_{t}A_{A}'+2iqr^{2}(\phi\bar{\phi}'-\phi' \bar{\phi})=0, \\
&A_{A}''+\dfrac{2iq}{r^{2}}(\phi \overline{D_{t}\phi}-D_{t}\phi\overline{\phi})=0, \\
&(r^{2}\phi')'-\dfrac{1}{r^{2}}(\partial_{t}^{2}\phi-iq\partial_{t}A_{A}\phi-2iqA_{A}\partial_{t}\phi -q^{2}A_{A}^{2}\phi)=0.
\ea
We then consider fluctuations around the classical background solutions, $A_{A}=A_{A}^{cl}(r)+\delta A_{A}(r,t),\ \phi=\phi^{cl}(r)+\delta \phi (r,t)$, with $\phi^{cl}$ taken to be real. It is convenient to introduce the linear combinations $\delta \phi_{R}=\delta \phi +\delta \overline{\phi}$ and $i\delta \phi_{I}=\delta \phi -\delta \overline{\phi}$. We then assume a homogeneous time dependence in the fluctuations as $\delta F=\delta F(r)e^{-i\omega t}$, where $\delta F=(\delta A_{A},\ \delta \phi,\ \delta \overline{\phi}$). The EOMs satisfied by the fluctuations are then given by
\ba\label{EFI14}
&{q}^{2}{{A^{cl}_{A}}}^{2}{\delta \phi_{R}} \left( r \right) +2i{ A^{cl}_{A}
}\omega {\delta \phi_{I}} \left( r \right) q+4{A^{cl}_{A}}{\phi^{cl}}{q}
^{2}{\delta A_{A}} \left( r \right) +{r}^{4}{
\delta \phi_{R}''} \left( r \right) +2{r}^{3}{\delta \phi_{R}'} \left( r
 \right) \nonumber  \\
&+{\omega}^{2}{\delta \phi_{R}} \left( r \right)=0,
 \\
&2{\phi^{cl\prime}}{q}{r}^{2}{\delta \phi_{I}} \left( r \right) -2q{
\phi^{cl}}{r}^{2}{\delta \phi_{I}'} \left( r \right) +i
\omega\,{\delta A_{A}'} \left( r \right)=0,  \nonumber \\
&{ {{A^{cl}}}^{2}{\delta \phi_{I}} \left( r \right) {q}^{2}+
  {\delta\phi_{I}''} \left( r \right)
  {r}^{4}+2{\delta\phi_{I}'} \left( r
 \right)   {r}^{3}-2i{A^{cl}}\omega {\delta \phi_{R}} \left( r
 \right) q-2i\omega{\delta A_{A}}q{\phi^{cl}}+{\omega}^{2}{\delta \phi_{I}}
 \left( r \right) } \nonumber \\
& =0, \nonumber \\
& {4{q}^{2}{A^{cl}}{\phi^{cl}}{\delta\phi_{R}} \left( r \right)
+4{q}^{2}{{\phi^{cl}}}^{2}{\delta A_{A}} \left( r \right) +2i\omega {
\delta \phi_{I}} \left( r \right) {\phi^{cl}}q-  {\delta A_{A}''} \left( r \right)   {r}^{2}}=0. \nonumber
\ea
These EOMs are not independent. It can be shown that the radial derivative of the second equation in \eqref{EFI14} is equal to a linear combination of the third equation and the fourth equation in \eqref{EFI14}. The number of integration constants hence is 5, due to the first order constraint.

In the remainder of this section we consider an infinitesimal $t_{hop}$ region, since the presence of  $t_{hop}$ changes the asymptotic expansion of perturbations.\footnote{In the presence of a nontrivial difference in the charge densities on both sites, i.e. at nontrivial $t_{hop}$, the asymptotic expansion of the background as well the perturbations changes to the more complicated form as in \protect\eqref{BIF314}. For large $t_{hop}$ an analytic approach is hence difficult. At leading order in $t_{hop}$ however, all the correction terms in \protect\eqref{BIF314} which are of higher order in $t_{hop}$ drop out, and we can proceed analytically.}
 In particular, the powers of $r$ in the asymptotic expansion for $\phi$ (\ref{BIF314}) are $t_{hop}$-independent, which will not be true for its perturbation, $\delta \phi_I$. These corrections at finite $t_{hop}$ also make it difficult to compute two point functions numerically, due to correction terms to the nonnormalizable modes, which turn out to be leading compared to the normalizable piece. In the following analysis, we will not employ an approximation where one field is taken to be a probe with respect to the other field. Rather, $\phi$ and $A_t$ are coupled when the sites have unequal charges and both fields enjoy both a homogeneous solution as well as inhomogeneous contributions. By studying the equations in the UV limit, $r \rightarrow \infty$, we perturbatively compute the subleading inhomogeneous contributions of both fields to each other. The small $t_{hop}$ analysis will  suffice to substantiate the statements that the phases inside the lobes of Fig.~\ref{fig:Lobe} are charged Mott insulators, while the phase outside does not have a gap of the order of the Coulomb parameter $U$, and hence can't be identified with a Mott insulator. The non-homogeneous phase will in particular show two almost zero modes, i.e. have a parametrically smaller gap. Note that we employ the same boundary conditions for the fluctuations that we used for the background configurations $A_A^{cl}$ and $\phi^{cl}$.

\subsection{Mott Insulator Phase}

In Mott insulator phase, $\phi^{cl}=t_{hop}$ and $A^{cl}=0$. The solutions $\delta A_{A}'$, $\delta \phi_{I}'$, and $\delta \phi_{R}$
 for finite hopping $t_{hop}$ are then given by
\ba\label{SOL05}
&\delta \phi_{I}'=\dfrac{C_{2}}{{r}^{\frac{5}{2}}} \mbox{BesselJ} \Big(-\alpha_{p},\dfrac{\omega}{r}\Big)+\dfrac{C_{3}}{{r}^{\frac{5}{2}}} \mbox{BesselY}\Big(-\alpha_{p},\dfrac{\omega}{r}\Big), \nonumber \\
&i\omega\delta A_{A}'={2qt_{hop}r^{2}\delta \phi_{I}'}, \quad \delta \phi_{R}=\delta \phi^{(1)}_{R}\cos \Big(\dfrac{\omega}{r}\Big) +\delta \phi^{(2)}_{R} \sin \Big(\dfrac{\omega}{r}\Big),
\ea
where $\alpha_{p}=\frac{1}{2}\sqrt{16q^{2}t_{hop}^{2}+1}$. $\delta \phi_{I}$ has the asymptotic behavior $\delta \phi_{I}\sim \varphi+ C_{3}r^{\alpha_{p}-\frac{3}{2}}$ ($\varphi$ is constant) near the boundary. The solution for $\delta\phi_I'$ is obtained by using the second equation in \eqref{EFI14} to transform the third equation in \eqref{EFI14} into a third order equation in $\delta\phi_I$. As seen in the $t_{hop}=0$ case, $\varphi$ and $C_{3}$ can be understood as the source term and VEV, respectively. Note that \eqref{SOL05} is obtained in the Mott insulator phase, and not in the inhomogeneous phase which leads to \eqref{BIF314}, and hence the powers in the falloffs of the fluctuations are different.

\begin{figure}[htbp]
  \begin{center}
   \includegraphics[height=5cm]{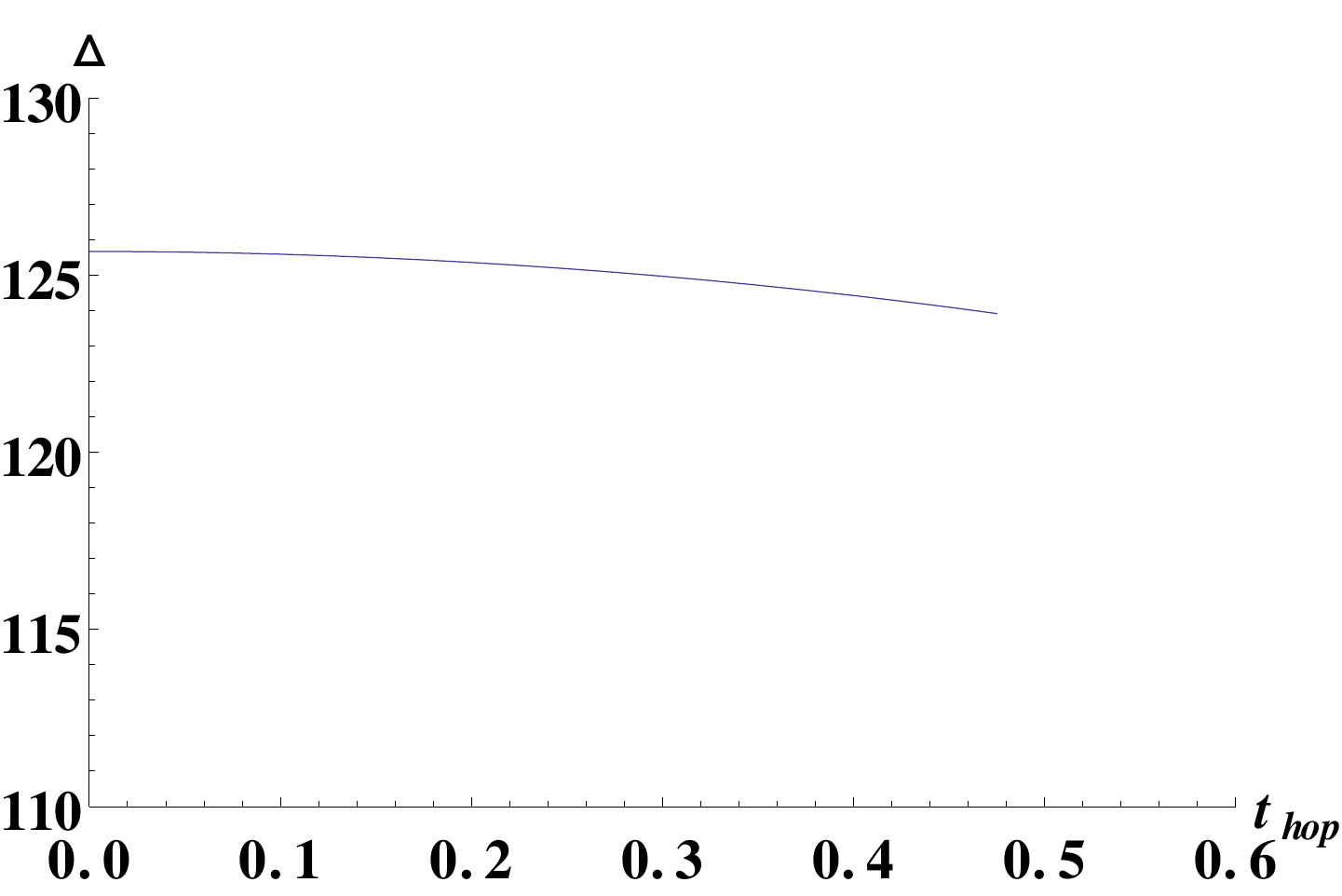}
  \caption{
  Gap of the first excitation as the function of $t_{hop}$. The plot does not depend on chemical potential $\mu_{b}$. The value $t_{hop,max}= 0.473$ is the largest value at which the Mott insulator/non-homogeneous phase transition happens for the effective potential in sec. 3, i.e. the value of $t_{hop}$ at the tip of the lobe. We observe that the gap in the Mott phase hardly changes with $t_{hop}$.
  }
    \label{fig:massMott}
  \end{center}
\end{figure}

For small hopping parameter ($\alpha_{p}\to\frac{1}{2}$), the Bessel functions and their integrals are replaced by $\cos$ and $\sin$ functions. The solutions are given by
\ba
&\delta A_{A}=-i\omega \varphi +O(t_{hop}),\quad \delta\phi_{I} =2qt_{hop}\varphi +\delta \phi^{(1)}_{I}\cos \Big(\dfrac{\omega}{r}\Big) +\dfrac{\delta \phi^{(2)}_{I}}{\omega} \sin \Big(\dfrac{\omega}{r}\Big),\nonumber  \\
&\delta \phi_{R}=\delta \phi^{(1)}_{R}\cos \Big(\dfrac{\omega}{r}\Big) +\dfrac{\delta \phi^{(2)}_{R}}{\omega} \sin \Big(\dfrac{\omega}{r}\Big), \label{PER06}
\ea
where $\varphi$ is constant.
Note that the number of integration constants is 5 consistent with EOMs \eqref{EFI14}. To derive the holographic two point functions we impose the IR boundary conditions $\delta A_{A}|_{r=r_{h}}=0$, $\delta \phi_{R}|_{r=r_{h}}=0$  and $\delta \phi_{I}|_{r=r_{h}}=0$, i.e. our boundary conditions allow neither the chemical potential nor the zero bifundamental VEV to vary.  This shows that $\varphi$ is negligible at $O(t_{hop})$ and
\ba
\delta \phi^{(2)}_{I,R}=-\omega \delta \phi^{(1)}_{I,R}\cot \Big(\dfrac{\omega}{r_{h}}\Big).
\ea
The Dirichlet boundary condition makes the differential operator describing the small fluctuations Hermitian and consequently all eigenfrequencies are real. This is expected in the Mott insulating phase.
Recall that the $AdS$ boundary expansion becomes $\delta \phi_{I,R}\sim \delta \phi^{(1)}_{I,R}+\delta \phi^{(2)}_{I,R}/r$.
The two point function of the operator dual to $\delta \phi_{I,R}$ is then given by
\ba
G_{I,R}=\omega \cot \Big(\dfrac{\omega}{r_{h}}\Big).
\ea
Thus, the real part of the two point function is the same as the imaginary part of it. These two point functions have a pole at $\omega =\pi r_{h }n_{o}$ ($n_{o}\ge 1$)~\cite{Erlich:2005qh}. However, we do not observe a peak at $\omega =0$, or low lying modes at very small $\omega$; we do not find zero modes or near-zero modes in the spectrum of the Mott insulator phase at least for small $t_{hop}$. We conclude that the Mott phase is gapped, with the gap of the order of the Coulomb repulsion,
\be\label{MottGapUVBCs}
\Delta = \pi r_h = \pi U\,.
\ee
When we take into account finite $t_{hop}$ corrections, the position of the peak is corrected in the imaginary part of the Green's function. We can numerically analyze this correction by finding the zero of the constant part of $\delta \phi_{I}$ under the above IR boundary conditions because this constant part is the non-normalizable mode of $\delta \phi_{I}$.
The gap of the excitations decreases only slightly as $t_{hop}$ is varied from zero to the maximal value at the tip of the lobe, $t_{hop,max}=0.473$ (see Fig. \ref{fig:massMott}). Hence, the mass gap of the excitations stays of the order of the Coulomb repulsion throughout the Mott phase.

In the Bose-Hubbard model~\cite{Markus12}, it is known that the low-lying excitations are described by the motion of a fundamental boson from a site to a neighboring site. To move a fundamental boson from a site to a neighboring lattice site costs energy $U$ because of the repulsive Coulomb force between the fundamental bosons. The mass gap obtained above is consistent in order of magnitude with the mass gap $\Delta= U$ of the Mott insulator phase in the actual Bose-Hubbard model.

\subsection{Non-homogeneous Phase}\label{sec:nonhompert}

\begin{figure}[htbp]
  \begin{center}
   \includegraphics[width=0.45\textwidth]{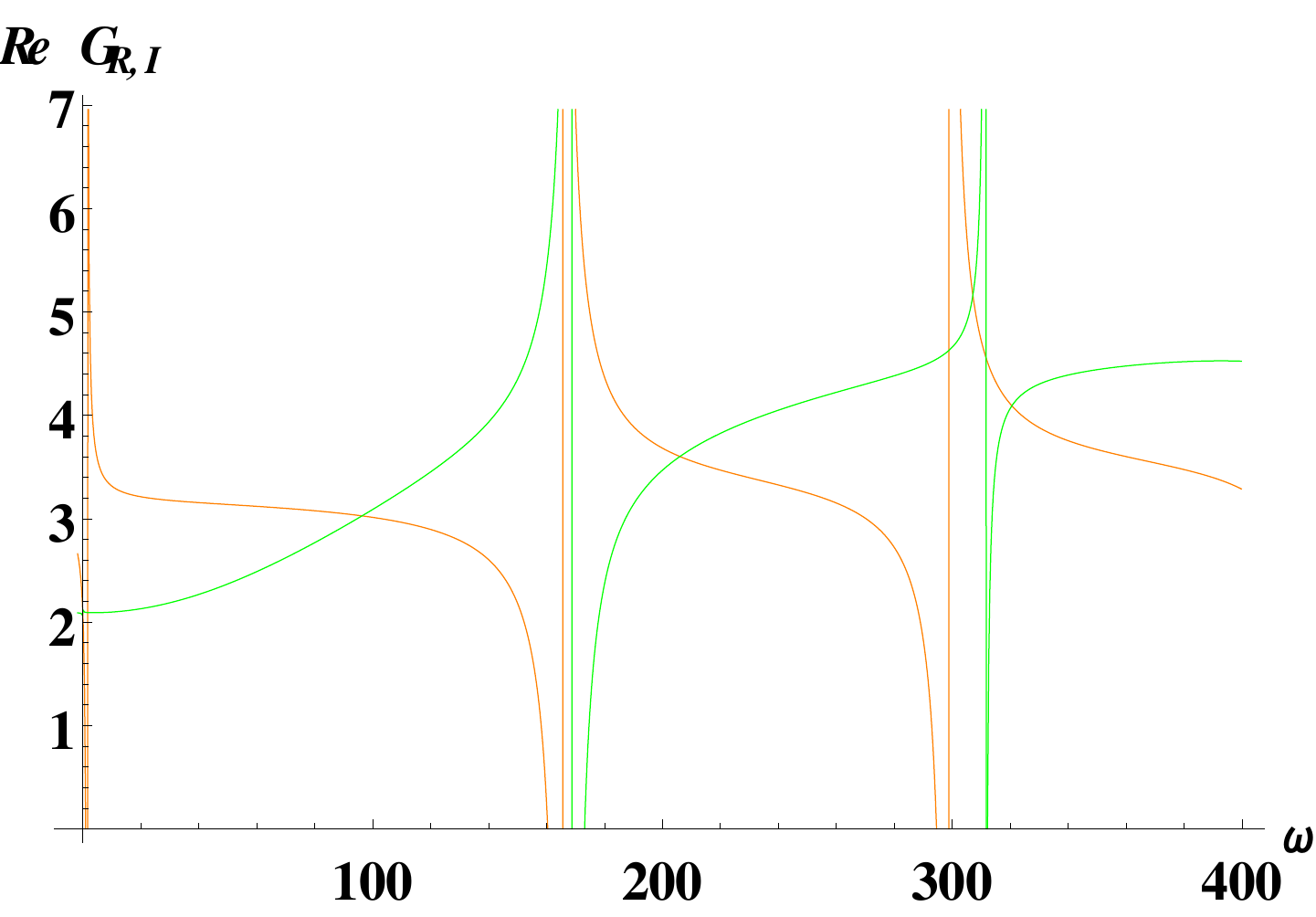}\hspace{0.5cm}
   \includegraphics[width=0.45\textwidth]{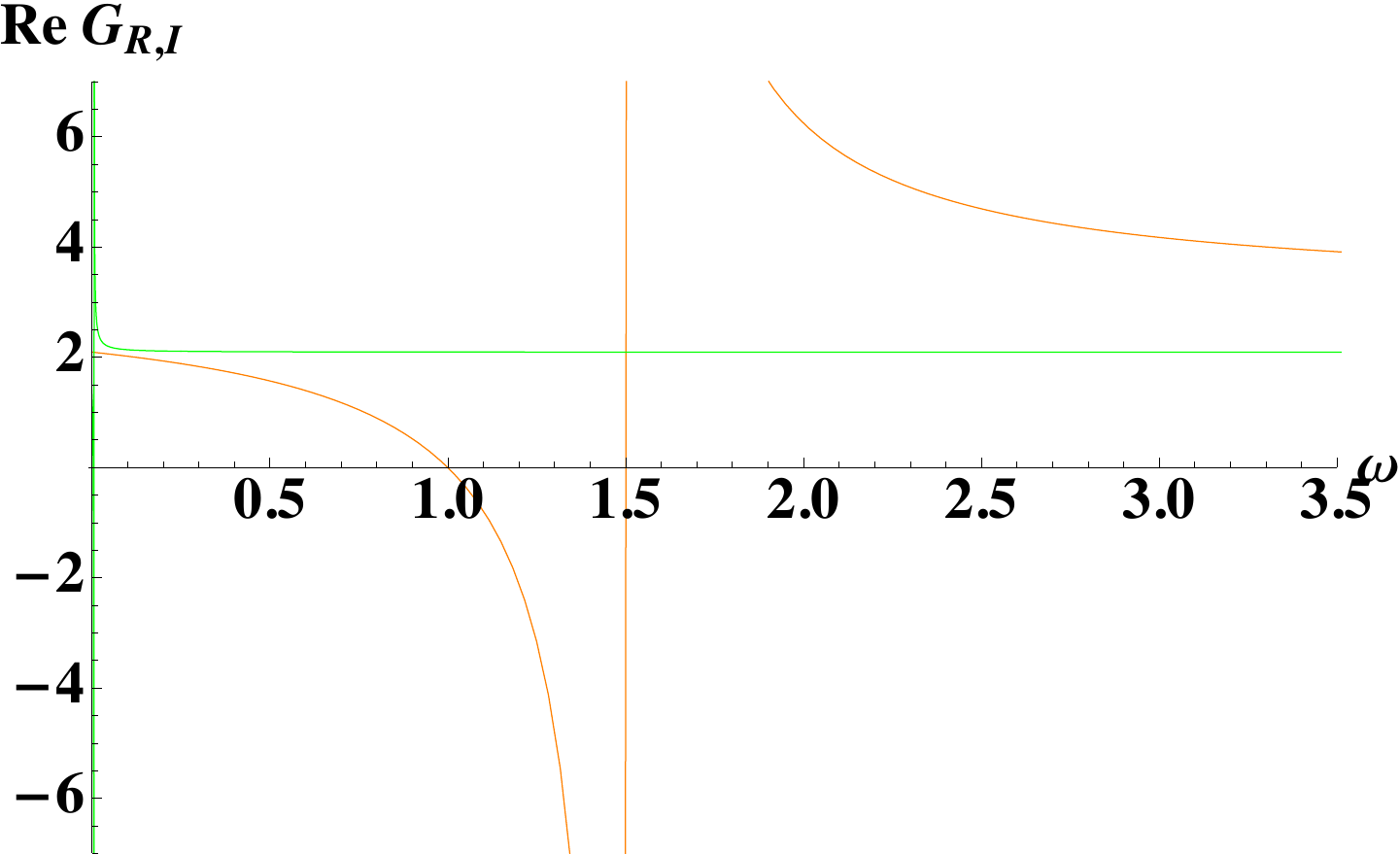}
  \caption{\textbf{Left Panel:} 
  Real parts of the two point functions $G_{R,I}$ dual to $\delta \phi_{I,R}$. Green and Orange curves show the two point functions dual to $\delta \phi_{R}$ and $\delta \phi_{I}$, respectively. $G_{R}$ (green curve) does not include zero modes but $G_{I}$ (orange curve) clearly includes an almost zero mode in its spectrum. \textbf{Right Panel:} Behavior of $G_{R,I}$ at small frequencies. We checked that no poles at zero frequency exist, but an almost zero mode both in $Re\, G_I$ (orange curve, at higher frequencies, already visible in the left panel) and $Re\, G_R$ (green curve, at very low frequencies). %The location of both poles varies with $t_{hop}$, but not in the usual way expected from spontaneous symmetry breaking, the Gell-Mann-Oakes-Renner \cite{GMOR} result $\omega_{\pi} \sim \sqrt{t_{hop}}$. 
  }
    \label{fig:ReG_RI}
  \end{center}
\end{figure}

\begin{figure}[htbp]
  \begin{center}
  \includegraphics[width=0.45\textwidth]{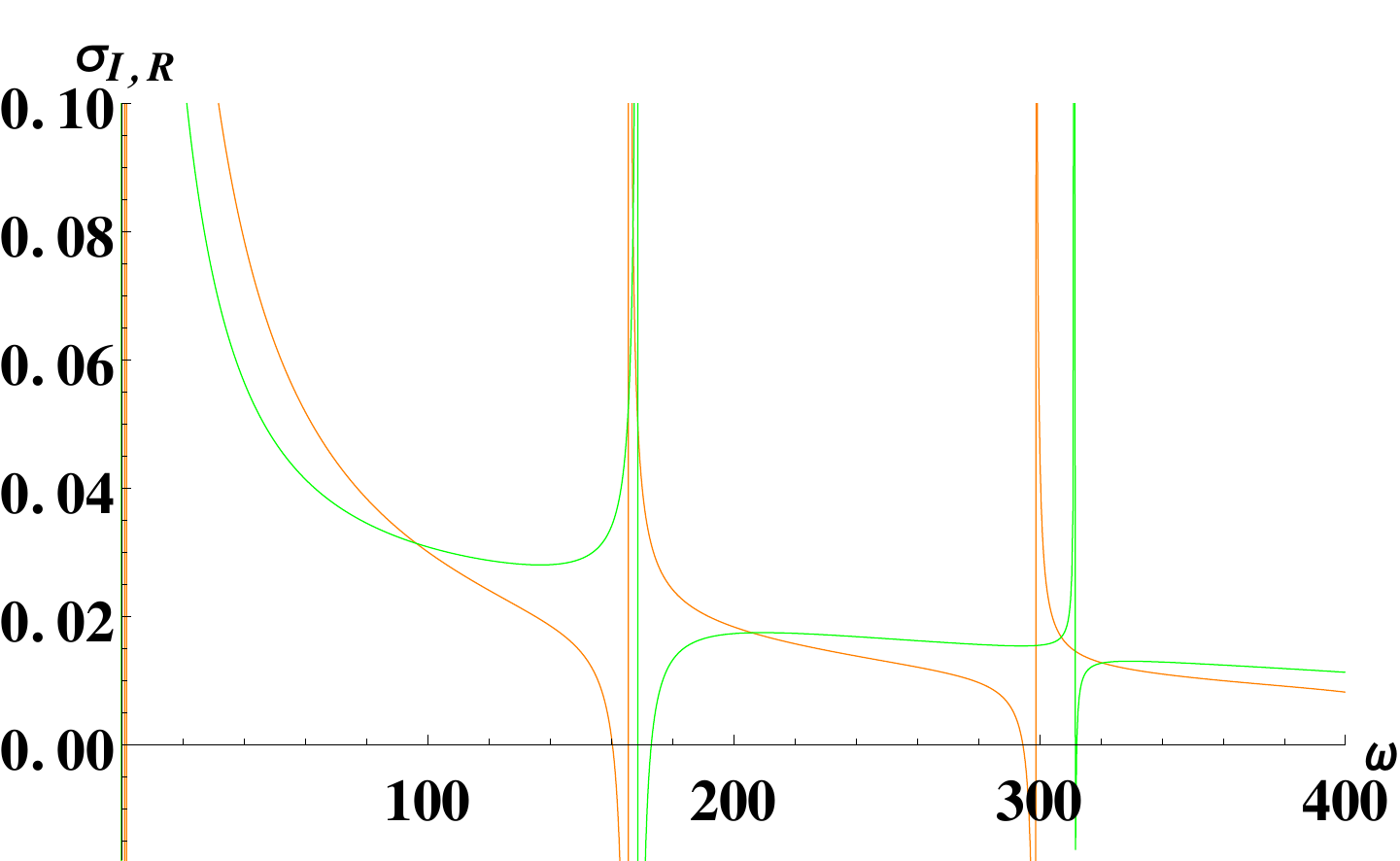}\hspace{0.5cm}
  \includegraphics[width=0.45\textwidth]{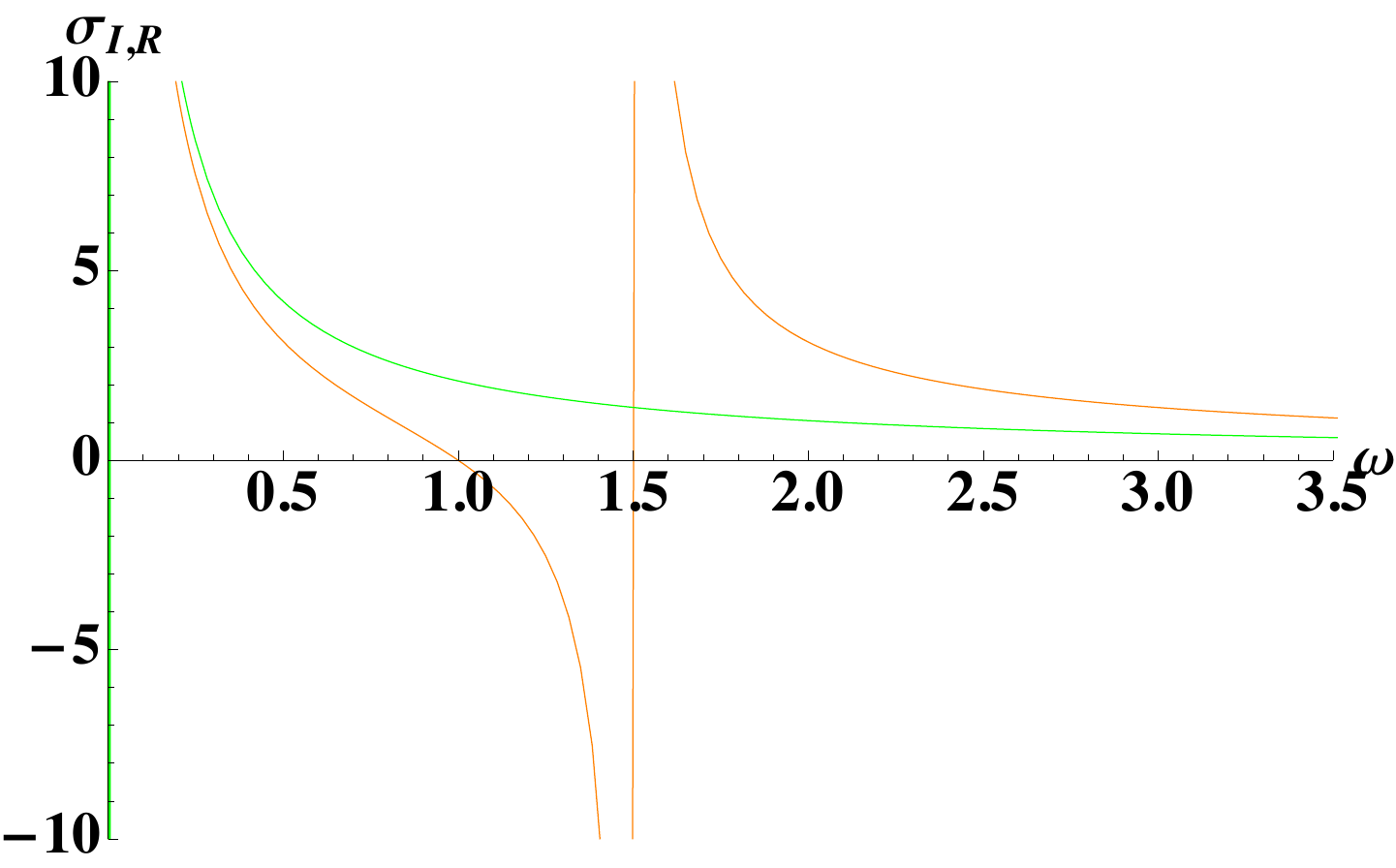}
  \caption{\textbf{Left Panel:} 
  Green and Orange curves show the ratio $G^{R,I}/\omega$ dual to $\delta \phi_{R}$ and $\delta \phi_{I}$ as a function of $\omega$, respectively. In Figure, $\sigma_{I,R}\equiv G^{I,R}/\omega$, respectively. Both $\sigma_{R,I}$ behave like $1/\omega$ at small $\omega$ larger than the energy of the almost zero modes found in fig.~\protect\ref{fig:ReG_RI}. %The peak near $\omega\sim 0$ does not almost move as $t_{hop}$ increases.  
  \textbf{Right Panel:} Behavior of the conductivities at small frequencies. The constant nonzero limit of $G^{R,I}$ as $\omega\rightarrow 0$ turns into a $1/\omega$ pole.
}
    \label{fig:Imsigma}
  \end{center}
\end{figure}

For small $t_{hop}$, the background in the non-homogeneous phase can approximated by
\be\label{smallthopBGapprox}
\phi^{cl}\sim t_{hop}r^{-\frac{2}{5}}\,,\quad A_{A}^{cl}\sim (\delta \rho) r\sim r\,.
\ee
This approximation, treating the bifundamental as a probe on top of the gauge field background, is possible as an inspection of the background equations of motion \eqref{EOE39} shows: For small $t_{hop}$ the gauge field coupling to the bifundamental is quadratic in $\phi$, and hence can be neglected.  
We consider  fluctuations around this approximated background in order to calculate the spectrum close to the cusps at $t_{hop}=0$ in the phase diagram. Our system of fluctuations is similar to the one in \cite{Mas:2008jz}, due to the coupling of scalar and longitudinal gauge modes.
Since they  do not admit analytic solutions, we solve the fluctuation EOMs numerically by the shooting method. We expand the fields at the hard wall as $\delta F=\sum_{n=0}\delta F^{(n)}(r-r_{h})^{n}$, where the parameters are fixed by 5 integration constants. The IR boundary conditions on the perturbations are chosen again to be Dirichlet boundary conditions $\delta A_{A}|_{r=r_{h}}=0$, $\delta \phi_{R}|_{r=r_{h}}=0$, and $\delta \phi_{I}|_{r=r_{h}}=0$.
Setting $r_{h}=40$, the expansion around the hard wall is then specified by 2 parameters $a,b$, $\delta \phi_{I} \sim a(r-r_{h})$, $\delta \phi_{R}\sim b(r-r_{h})$, and $\delta A_{A}\sim \delta F^{(1)}(a)(r-r_{h})$, where $\delta F^{(1)}(a)$ is given by $\delta F^{(1)}(a) = i t_{hop} a/\omega$. One can fix one of these constants to be 1 ($a=1$ for example) by rescaling the perturbations. For small $t_{hop}$, the asymptotic expansion at the AdS boundary becomes
\ba
&\delta\phi_{R}\sim (\phi_{R}^{as}-\frac{24}{5}t_{hop}A^{as}\log (r))r^{-\frac{2}{5}}+\phi_{R}^{(2)as}r^{-\frac{3}{5}}+\dots , \nonumber  \\
&\delta\phi_{I} \sim \phi_{I}^{as}r^{-\frac{2}{5}}+\phi_{I}^{(2)as}r^{-\frac{3}{5}}+\dots , \nonumber \\
&\delta A_{A}\sim A^{as}r+\dots  \label{ASY011}
\ea
We then fix the remaining constant $b$ by the boundary condition $A^{as}=0$. Note that this boundary condition does not depend on the frequency $\omega$. The holographic two point functions are then computed as
\be
G_{R,I}=-\dfrac{\phi_{R,I}^{(2)as}}{\phi_{R,I}^{as}}.
\ee
The two point functions are plotted in Fig.~\ref{fig:ReG_RI} numerically. We do not observe a peak at $\omega =0$ for $G_{R}$ but do observe a peak near $\omega = 0$ for both $G_R$ and $G_{I}$ (c.f. the right panel of fig.~\ref{fig:ReG_RI}). While both modes appear at frequencies much smaller than the Coulomb gap set by $r_h$, the mode in $G_R$ (green curve) appears at much smaller, possibly even parametrically smaller, frequency compared to the mode in $G_I$ (orange curve). %The existence of a zero mode in the bifundamental correlator $G_I$ would imply a Goldstone mode in the non-homogeneous phase, consistent with a possible interpretation as a  superfluid phase. In order to decide whether our mode is an 
%Since the background considered in this section has a small explicitly symmetry breaking source ($t_{hop}$) switched on, we would expect the Goldstone mode to become slightly massive due to the mixing of explicit and spontaneous symmetry breaking \cite{GMOR}. We however find that the almost zero modes both do not change its position in the usual Gell-Mann-Oakes-Renner \cite{GMOR} like fashion with $\omega_p\sim \sqrt{t_{hop}}$ as $t_{hop}$ is varied, suggesting that both are not actual Goldstone modes. Our numerical results instead indicate a change of the position of the modes with $t_{hop}$ as
%%
%\begin{eqnarray}\label{GMORfail}\nonumber
%\omega_{p,G_R} &=& 8\cdot10^{-5}-0.00207 \sqrt{t_{hop}} + 0.01965 t_{hop} - 0.08667 t_{hop}^{3/2} + 0.74470 t_{hop}^2 - 0.30506 t_{hop}^3 +\dots\,,\\
%\omega_{p,G_I} &=& 1.49826 + 6\cdot10^{-5} \sqrt{t_{hop}} - 37\cdot10^{-5} t_{hop} +\dots\,,
%\end{eqnarray}
%for the mode in $G_R$ and $G_I$, respectively.
%
%We will comment on ways to introduce superfluidity and hence generate actual spontaneous symmetry breaking in sec.~\ref{sec:Discussion}.

{The ratio $G_{R,I}/\omega$ is plotted in Fig. \ref{fig:Imsigma} as a function of $\omega$. In Fig. \ref{fig:Imsigma}, $\sigma_{I,R}\equiv G_{I,R}/\omega$. Both $\sigma_{R,I}$ behave like $1/\omega$ at small $\omega$ below the respective gaps, c.f. the right panel in fig.~\ref{fig:Imsigma}. In particular, $\sigma_{I}$ should be related with the conductivity of the Bose-Hubbard model because the current in the Bose-Hubbard model is defined by
\ba
\label{current}
J=t_{hop}\mbox{Im}(b_{i}^{\dagger}b_{j})= t_{hop}\phi_{I}^{(2)}
\ea
Note that the creation and annihilation operators are dimensionless, while $t$ has dimension of energy, so $J$ has the usual dimension 1. 

The identification \eqref{current} can easily be derived from a conservation equation. Note that charge conservation, as usual, should imply the following continuity equation:
\ba
\dot{\rho}_{(l)} = J_{l-1} - J_{l}. \label{continuity}
\ea
The right hand side is (minus) the discretized spatial derivative of the current at unit lattice spacing. Note that no lattice spacing appears here, as the lattice spacing $a_{1}$ in our lattice is set by $t_{hop} \sim a_{1}^{-1}$. We can use this relation to identify $J_l$. For time dependent fields, we can look at the $r$ component of Maxwell's equations which we so far neglected, as it is automatically solved for static configurations. Note that below we set $q=1$ by a rescaling of the bifundamental fluctuations.\footnote{In a more careful treatment the charge of the bifundamental under the axial gauge field combination would appear here, which, if we want to retain the standard normalization of the Maxwell term, will be related to the charge of the bifundamental at each site by $q_A = \sqrt{2} q$.} This EOM reads
\ba
\partial_t (A_t^{(l)})' = -j^r
\ea
where $j_r$ is the bulk current associated to the scalar fields. To meaningfully talk about a spatial current we should consider a multi-side model with bi-fundamentals $\phi^{l}$ connecting the $l$-th and $(l+1)-th$ site. In this case
\ba
j^r = i r^2  \left ( \left [ \phi^{l,*} \partial_r \phi^{l} - \phi^l \partial_r \phi^{l,*} \right ] -
\left [ \phi^{l-1,*} \partial_r \phi^{l-1} - \phi^{l-1} \partial_r \phi^{l-1,*} \right ] \right )
\ea
Keeping only the leading order $r$ terms in the near boundary expansion of the fields as in \eqref{BIF314} this EOM reads
\ba \label{finaleom} \dot{\rho}_{(l)} = \Im ( t_{hop,l-1} \varphi_{l-1} ) - \Im (t_{hop,l} \varphi_l). \ea
In this expression we have different leading behaviors for the scalar fields on the various sites, but the version of the Hubbard model we are considering has all $t_{hop,j}$ be equal to the same $t_{hop}$, which we further can chose to be real. With this our equation of motion \eqref{finaleom} can be compared to the continuity equation \eqref{continuity} to directly give \eqref{current}.

\section{Generalization to the $n$-site model}

So far, we employed the two-site model mostly for computational simplicity, and found its physics to be rather similar to the Bose-Hubbard model. In this section, we briefly introduce the generalization of our model to the $n$-site model.
The action of a model with $n$ sites is given by $S=S_{kin}+S_{matter}$ as
\ba
&S_{kin}^{(n)}=\sum_{k=1}^n\int d^2x\sqrt{-g}\Big(-\dfrac{1}{2}F^2_{(k)}\Big), \\
&S_{matter}^{(n)}=-\sum_{k=1}^n\int d^2x\sqrt{-g}|D_{(k)}\phi_k|^2-\sum_{k=1}^n\int dtr_h\Lambda (|\phi_k|^2+w^2)^2,
\ea
where $F_{(k)}^2=F_{(k)\mu\nu}F^{\mu\nu}_{(k)}/2$ and $D_{(l)}=\partial_{\mu}-iqA_{\mu}^{(l)}+iqA_{\mu}^{(l+1)}$. In the summation, if we consider a chain model, $n+1$ is identified with $1$. Other summations over different spatial lattices (triangle, honeycome, Kagome etc.) are straightforward to introduce. We can take the following linear combination to extract the diagonal gauge field $V_{\mu}$ as
\ba
V_{\mu}=\sum^n_{l=1}A^{(l)}_{\mu},\quad A_{A\mu}^{(l)}=A^{(l)}_{\mu}-A^{(l+1)}_{\mu}.
\ea
The Maxwell kinetic term is rewritten as
\ba
&-\dfrac{1}{2}\sum^n_{k=1}F^{2}_{(k)}=-\dfrac{1}{2n}\Big[F_V^2+\sum_{k=1}^{n-1}F_{A(k)}^2+\sum_{k=1}^{n-2}(F_{A(k)}+F_{A(k+1)})^2 \nonumber \\
&+\sum_{k=1}^{n-3}(F_{A(k)}+F_{A(k+1)}+F_{A(k+2)})^2+\dots+(\sum_{l=1}^nF_{A(l)})^2\Big].
\ea
The covariant derivative which appears in $S_{matter}$ can be rewritten as $D_{(l)}=\partial_{\mu}-iqA_{A\mu}^{(l)}$ for $l=1,\dots ,n-1$ and $D^{(n)}=\partial_{\mu}-iq\sum_{l=1}^{n-1}A^{(l)}_{A\mu}$ by using the gauge fields $A_{A\mu}^{(l)}$. The Maxwell term of the diagonal gauge field $V_{\mu}$ then decouples from the remaining part of the action as
\ba
S^{(n)}_{kin}+S_{matter}^{(n)}\equiv -\dfrac{1}{2n}\int d^2x\sqrt{-g}F^2_{V}+K[A_{A}^{(1)},A_A^{(2)},\dots,A_A^{(n-1)},\phi_l,\dots,\phi_n].
\ea
This implies that the free energy is of the form $F=\mu\sum_{i}\rho_{(i)}+E(\rho_{(i)},t_{hop})$ like \eqref{FRE315}. For the chain the physics will be similar to the two-site model; in particular, the level-changing phase transitions will work in the same way, and the phase diagram will be qualitatively unchanged. It would be interesting to explore the phase structure of this model for different lattice configurations and/or beyond-nearest-neighbor hoppings. 

\section{Discussion}\label{sec:Discussion}

In this work we have analyzed a holographic dual of the Bose-Hubbard model based on U(1) gauge fields localized on gapped $AdS_2$ hard wall space-times, which are connected to each other by bifundamentals charged under the respective gauge groups. We have shown that the model admits a good one-to-one holographic dictionary with the operators and parameters showing up in the Bose-Hubbard model, that a model based on two sites already reproduces the lobe-like phase structure in the chemical potential - hopping parameter ($\mu_b-t_{hop}$) plane, that the Mott insulating states have a natural excitation gap of the order of the Coulomb repulsion parameter, and that the transition to the inhomogeneous phase at the cusp points at zero hopping where the lobes meet is second order.

Our holographic model exhibits several differences from the Bose-Hubbard model: Except at the cusp points, the transition to the inhomogeneous phase is generically first order. %and no Goldstone modes associated with the spontaneous breaking of the difference U(1) gauge groups are present in the inhomogeneous phase. 
In the excitation spectrum we find two near-zero modes at unnaturally small frequency appearing in the inhomogeneous phase near the cusp points. A priori, these modes could be connected to the spontaneous breaking of the difference U(1) gauge group in our two-site model. A preliminary analysis showed that these near zero modes change their position with varying hopping parameter, but whether they show the correct variation for a Nambu-Goldstone mode \cite{GMOR} can only be decided by a more precise numerical analysis. This question and also whether these conclusions continue to hold in other parts of the phase diagram will be the topic of a future, more complete investigation of the fluctuation spectrum \cite{WIP}. Finally, the overall vector U(1) in our model is not broken by the hopping, while it is in the condensed state of the Bose-Hubbard model \cite{Fisher:1989zza}.

In view of these differences to the Bose-Hubbard model, the two most interesting questions for future work will be: how to achieve a continuous phase transition between Mott and inhomogeneous phases everywhere along the phase boundaries, and how to achieve superfluidity in the inhomogeneous phase. A continuous phase transition is  generically expected in holographic models with spontaneous breaking of U(1) symmetries  in the bulk \cite{Nishioka:2009zj}.  A hint to the issue is the $AdS_2$ hard wall geometry we are using, which is not a solution to Einstein's equations, so the first order nature of the phase transition may be an artifact of this shortcoming. {An obvious improvement would be to use an $AdS_2$ hard wall-like geometry as is typically induced on the worldvolume of effectively two-dimensional probe branes embedded into higher dimensional $AdS$ solitons \cite{Horowitz:1998ha}.} In these geometries, the radial direction would cap off smoothly, and hence smoothen out the phase structure. Furthermore, such a bottom-up model would be more easily connected to top-down constructions of the bosonic and fermionic Hubbard models (see below).

Even in {the hard wall-like geometry induced on the probe brane}, however, the transition may still be first order, due to the transition taking place in the presence of a finite source term switched on, the hopping parameter. In this case we will need to introduce additional superfluid order parameters which should condense spontaneously, in order to achieve actual superfluidity in the inhomogeneous phase. There are basically two options: either we can couple fundamental scalars to the U(1) gauge field at each site, or introduce additional bifundamentals. In the former case the U(1) will break spontaneously if the local charge density at a particular site exceeds a critical value given by the charge and mass of the fundamental at that site, while in the latter the different charge density between the two sites to which the bifundamental is connected will be important. Specializing to the two-site model, we for example can break the vector U(1) spontaneously by introducing an additional fundamental at either of the sites. On the other hand, the two-site construction used in the main part of this paper, where we set the normalizable (i.e. in a sense spontaneously generated) part of the kinetic energy VEV to zero by our UV boundary conditions, could easily be amended by introducing a second bifundamental with the same charge as the first one (but maybe different mass), which again connects both sites. In this case however we would require this second bifundamental to condense with zero source term, i.e. not switch on a hopping parameter for it. The combined dynamics of this extended two-site model would then exhibit spontaneous breaking from the second bifundamental, while an explicit hopping VEV would be generated from the ``hopping bifundamental". 
We are planning to present results on these different possibilities, as well as on other improvements of the model, in a follow-up work \cite{WIP}.

In this work we have mostly focused on a simple bottom-up construction. Here we would like to outline how to construct a top-down version of our model using the $AdS_5$ soliton \cite{Horowitz:1998ha}. We introduce a probe D5-brane on the $AdS_5$ soliton times $S^5$ \cite{Horowitz:1998ha} without considering its back-reaction.\footnote{{Focusing on the 3-5 string modes where the ground state of the massless mode is obtained from R-sectors, such a state is given by the fundamental fermions $\chi_{ia}$ of $U(N)$ gauge symmetry on a site. So, this D3-D5 model seems to be a good holographic dual to the Fermi-Hubbard model. However, note that in such top-down constructions one is usually forced to work in the strict large $N$ limit, in which we do not expect as many differences between the fundamental bosons and the fundamental fermions, since large $N$ numbers of particles can occupy states of the same energy, as is the case in boson statistics. In other words, there is no restriction from the Pauli exclusion principle at large $N$. We do expect, however, at least one crucial difference between bosons and fermions, even in the strict large $N$ limit: in the presence of fermionic anomalies, the probe action of the dual D-branes will contain a Wess-Zumino (WZ) term. This term will be absent for the duals to bosonic fields, which do not contain anomalies. See \cite{Yamaguchi:2006tq, Drukker:2005kx, Hartnoll:2006ib}.}}
Recall that the D5 probe branes can not end at the tip of the soliton (hard wall) and the D5-brane has to come back at a turning point like in holographic QCD \cite{Sakai:2004cn} (see also~\cite{Kachru:2009xf,Kachru:2010dk}). If we do this in the internal soliton directions, the brane may smoothly cap off before hitting the soliton. However, we can also introduce a D7-brane filling the 2+1 dimensional uncompactified directions of the cigar and wrapping the whole $S^5$. When we have a D7-brane sitting at the tip of the $AdS$ soliton, the D5-brane can end on this D7-brane. Such a D7-brane has been identified as the holographic level-rank dual to Chern-Simons theory in~\cite{Fujita:2009kw}. In summary, these two world-volume theories seem to be good top-down constructions. We can expect the fundamental fermions corresponding to the 3-5 string modes to have the following low energy effective action at each site:
\ba
\int dt \Big(i\chi^{\dagger ia}\partial_t\chi_{ia}+\chi^{ia\dagger}(A_0+\phi_{D3})_i{}^j\chi_{ja}+\chi^{ia \dagger}[\tilde{A}_0]_a{}^b\chi_{bi}\Big)+S_{\mathcal{N}=4}+S_{extra}, \label{ACT01v2}
\ea
where $\phi_{D3}$ is the transverse scalar, $A_0$ is the D3 gauge field, and $\tilde{A}_0$ is the D5 gauge field. Considering the background of the D5-brane gauge field $[\tilde{A}_0]_{a}{}^a=\mu_{a}$ and $[\tilde{A}_0]_{a}{}^b=t_{ab}$ $(a\neq b)$, we have the following hopping term and chemical potential from the third term in \eqref{ACT01v2}:
\ba
\int dt\sum_{a=1\dots N_f}(i\chi^{\dagger ia}\partial_t\chi_{ia}+ \mu_a\chi_{ia}^{\dagger}\chi_{ia})+\sum_{a\neq b}(t_{ab}\chi^{i,a \dagger}\chi_{i,b}+c.c. )+\dots. \label{HOP03v2}
\ea
This action is similar to semi-holographic fermions \cite{Faulkner:2010tq,Jensen:2011su} in the absence of the second fermionic operator. The holographic dual to this field theory would then be a stack of D5 branes, possibly ending on D7 branes, which are separated from each other to reside on the different sites of our model. The nonabelian D5 brane gauge symmetry is higgsed to the U(1) subgroups in this process, and the off-diagonal components of the nonabelian D5 gauge field will become the bifundamentals in our bottom-up construction \cite{Kachru:2009xf,Kachru:2010dk}.

The construction with D7 branes at the soliton tip seems to have another advantage: In our hard wall model of the bottom-up construction, we added a potential in the IR boundary. In the top-down construction, the interactions between the different D5-branes will be deformed by the Chern-Simons terms on the hard wall. Moreover, the action of 5-7 string modes will introduce additional interactions at the D5/D7 intersections as pointed out in a slightly different set-up in Ref.~\cite{Erdmenger:2013dpa}. In that reference, $D7$-branes are suspended from the $AdS$ boundary, whereas in our model, in order to serve as an effective IR boundary to the D5 branes, the D7-brane has to sit at the tip of the $AdS$ soliton. This can be achieved by having defect D7 branes falling into the bulk from the boundary which then must, by charge conservation, bend back to the boundary by reversing their orientation, similar to what happens to the D8 branes in \cite{Sakai:2004cn}. The bent-back D7 brane will hence be a $\overline{D7}$ brane, and in the limit of large separations of the boundary defects, the D7 brane will sit at the tip of the cigar for a long distance. It will have an effective description as an infinitely extended D7 brane parallel to the AdS boundary, at which the D5 branes can then end. Such a top-down construction would be very useful, since we are not required to tune free parameters. We are planning to analyze these different top-down constructions in detail in the near future \cite{WIP}.

Generalizing our model to higher-dimensional Bose and Fermi Hubbard models  would be interesting because the higher-dimensional Bose and Fermi Hubbard models are more difficult to analyze using field theoretic or numerical techniques. In higher dimensions,  Monte Carlo simulations are typically used to analyze the ground and thermal states, but suffer, in the fermionic case, from a sign problem. Different lattices (tetrahedral, triangular, Kagome, etc.) lead to vastly different physics such as spin-charge fractionalization, frustration, quantum spin ices, spin liquids, etc.. It will be very interesting to investigate the possible phases of higher dimensional lattices in this model in future work. It would also be interesting to apply our model to disordered systems in one or higher dimensions, by e.g. randomizing the chemical potentials, the hopping parameter, or other parameters (such as the bifundamental mass \cite{Arias:2012an} or parameters in the IR potential). In disordered systems in the bosonic case, the Bose glass-phase appears in the phase structure between the insulating and superfluid phases. A Bose glass is characterized by a vanishing gap and finite compressibility, but it is an insulator because localization occurs due to the random potential. In the real Bose-Hubbard model, the phase transition to the superfluid phase is known to occur only from the Bose-glass phase \cite{Fisher:1989zza}. Disorder was introduced in AdS/CFT in Refs.~\cite{Hartnoll:2008hs,Fujita:2008rs,Arean1,Arean2,Hartnoll1,Hartnoll2,Lucas:2014zea}. For Gaussian disorder, the free energy of the disordered system can in particular be evaluated by introducing replica fields and by averaging over disorder: $F=-\overline{\log Z}$ where $\log Z=(Z^{n}-1)/n$ as $n\to 0$ \cite{Fujita:2008rs}.

\paragraph*{Acknowledgement:}
Special thanks to Shamit Kachru for collaboration during early stages of this work, as well as to Edward Witten for pointing out to us the relation between monopole couplings and the quantization of overall charge in generic field theories.
We would like to thank T. Azeyanagi, J. Bhattacharya, S. Das, M. Hanada, S. He, M. Kaminski, S. Minwalla,  T. Nishioka, M. Shigemori,  S. Sugimoto, T. Takayanagi, M. Tezuka, and A. Trombettoni for helpful discussions and comments. M. F. is in part supported by JSPS Postdoctoral Fellowship and partly by JSPS Grant-in-Aid for JSPS Fellows No. 25-4348. The work of R.M. was supported by World Premier International Research Center Initiative (WPI), MEXT, Japan. S.M.H. is supported by the Harvard University Lawrence Golub Fellowship in the Physical Sciences. The work of A. K. is in part supported by the U.S. Department of Energy under Grant number DE-SC0011637. N. P. is supported by a Stanford Humanities and Sciences Fellowship and an NSF Graduate Research Fellowship.

\appendix

\section{Mixed Neumann boundary condition}\label{App:MixedNeumann}

In this appendix, we consider the mixed Neumann IR boundary conditions arising from the variation of the action instead of the charge quantization and zero VEV boundary conditions of the bi-fundamentals that we imposed in the UV so far (which correspond to free boundary conditions in the IR). We find that this boundary condition naturally generates a VEV for the bi-fundamentals in both the Mott insulator and the non-homogeneous phase.

\subsection{Homogeneous Mott Insulator}
The homogeneous phase is defined by the condition $A_t^{(1)}=A_t^{(2)}$. The EOM of the fields $\phi$ and $\ A_t^{(l)}$ are then diagonalized. The EOM of the fields \eqref{EOE39} are solved analytically as
\ba\label{SOP310ap}
\phi=t_{hop}+\dfrac{\varphi_0}{r},\quad A_t^{(l)}=\mu+\rho_{(l)}r,
\ea
where $l=1,2$. The coefficient $t_{hop}$ and the coefficient of the normalizable mode $\varphi^{Mott}\equiv -\varphi_0$ are identified with the hopping parameter and vacuum expectation value (VEV) of the bi-local field $b_{i}^{\dagger}b_{j}$ in Bose-Hubbard model, respectively. Note that the minus sign in front of $\varphi_{0}$ appears in the formula of VEV (see~\cite{Mateos:2006nu,Mateos:2007vn}).

\begin{figure}[htbp]
  \begin{center}
   \includegraphics[height=7cm]{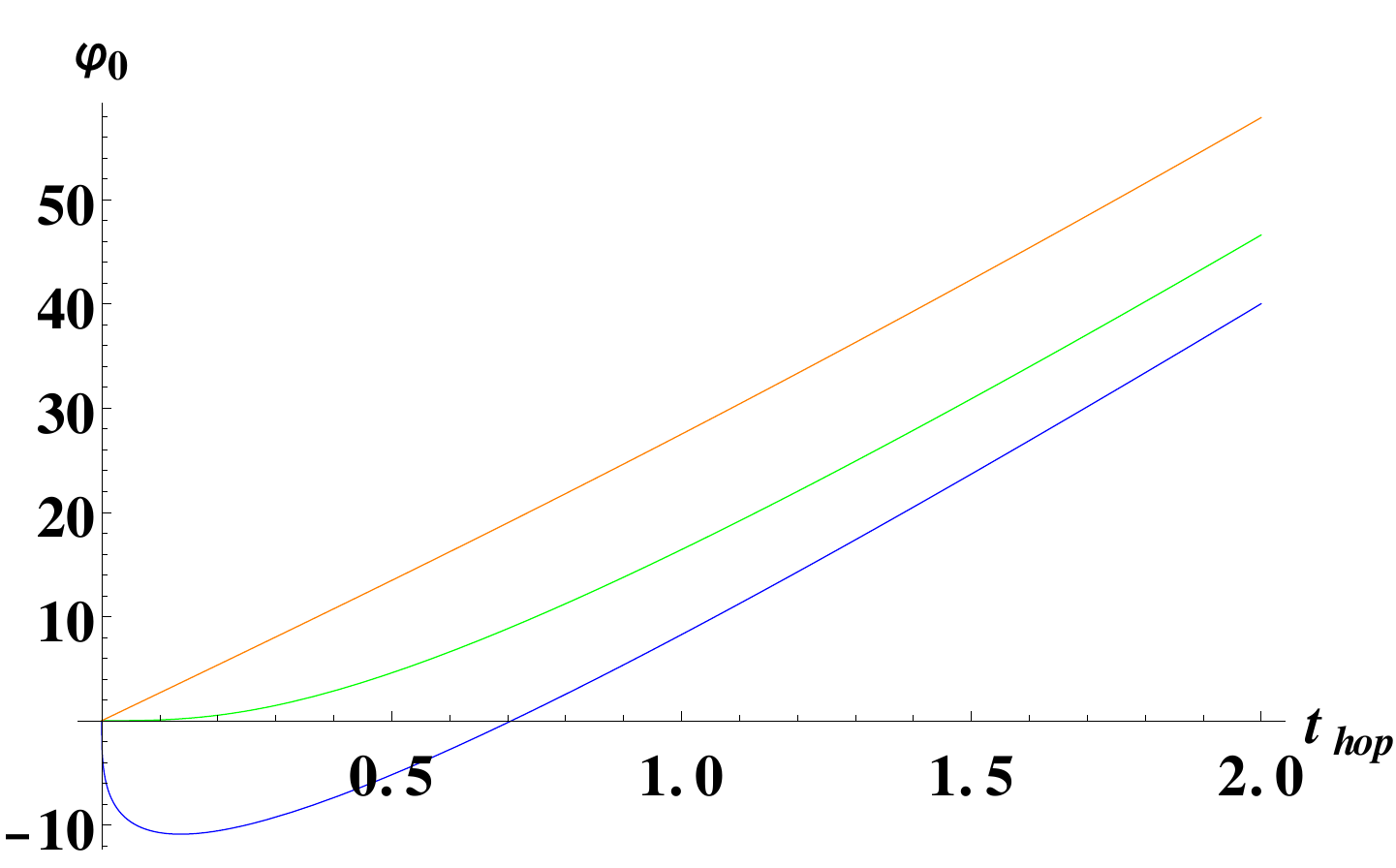}
  \caption{
  VEV $\varphi^{Mott}(=-\varphi_{0})$ as the function of $t_{hop}$ ($\Lambda=1$). Blue line: $w^{2}=-1/2$. Green line: $w^{2}=0$. Orange line: $w^{2}=1$. When $w^{2}<0$, $\varphi^{Mott}$ changes the sign.
  }
    \label{fig:vphi}
  \end{center}
\end{figure}

The boundary term from varying the action is required to vanish at the IR wall, giving rise to the mixed Neumann boundary condition  (see also~\cite{Csaki:2003dt,Csaki:2003zu})
\ba\label{IRB16ap}
-r_{h}\phi'+2\Lambda \phi (\abs{\phi}^{2}+w^{2})=0.
\ea
The above boundary condition is a generalized version (due to the IR potential) of a class of modified boundary conditions which can e.g. describe a metal/insulator phase transition~\cite{Nishioka:2009zj,Horowitz:2010jq}.
The condition \eqref{IRB16ap} can be solved for the VEV $\varphi_{0}$ ,
\ba
&\varphi_{0}=\Big(-r_{h}t_{hop}-\dfrac{r_{h}(1+2\Lambda w^{2})}{6^{1/3}\sqrt{\Lambda}Q_{1/3}}+\dfrac{r_{h}Q_{1/3}}{6^{2/3}\sqrt{\Lambda}}, \label{SOP17} \\
& -r_{h}t_{hop}+\dfrac{(1+\sqrt{3}i)r_{h}(1+2\Lambda w^{2})}{2\cdot 6^{1/3}\sqrt{\Lambda}Q_{1/3}}+\dfrac{(-1+\sqrt{3}i)r_{h}Q_{1/3}}{2\cdot 6^{2/3}\sqrt{\Lambda}}, \nonumber \\
& -r_{h}t_{hop}+\dfrac{(1-\sqrt{3}i)r_{h}(1+2\Lambda w^{2})}{2\cdot 6^{1/3}\sqrt{\Lambda}Q_{1/3}}+\dfrac{(-1-\sqrt{3}i)r_{h}Q_{1/3}}{2\cdot 6^{2/3}\sqrt{\Lambda}}\Big), \nonumber
\ea
where $Q_{1/3}=(9\sqrt{\Lambda}t_{hop}+\sqrt{6+72\Lambda^{2}w^{2}+48\Lambda^{3}w^{3}+9\Lambda (9t_{hop}^{2}+4w^{2})})^{\frac{1}{3}}$.
 We can show that the real solution \eqref{SOP17} of the (generally complex) three solutions of \eqref{IRB16ap} minimizes the on-shell action below (free energy). Moreover, the two complex solutions are non-zero at $t_{hop}=0$ and are hence not preferred (we already chose a gauge in which $\phi$ is real). We plot the VEV $\varphi^{Mott}(=-\varphi_{0})$ of the real solution \eqref{SOP17} as the function of $t_{hop}$ in Fig. \ref{fig:vphi}. When $w^{2}<0$, $\varphi_{0}$  changes  sign, which is not physically preferred either - we would like the VEV of the kinetic energy operator to be positive for positive hopping parameter.  We hence choose a positive mass $w^{2}$ in what follows.

In the homogeneous phase, $S_{kin}$ is finite and additional counter-terms are not needed. The free energy is given by
\ba
&F_{Mott}=-(S_{kin}+S_{cut}+S_{matter})/\beta  \nonumber \\
&=2\mu\rho_{(1)}+r_h\rho_{(1)}^2 +\dfrac{\abs{\varphi_{0}}^{2}}{r_{h}}+r_h\Lambda (\abs{\phi(r_{h})}^2+w^2)^2.
\ea

\subsection{Non-homogeneous Mixed State}
For the non-homogeneous case $A_t^{(1)}\neq A_t^{(2)}$ there is no analytic solution to the EOM \eqref{EOMinhom} in general. We rely on numerical methods to solve the EOM \eqref{EOMinhom}.

We obtain the following asymptotic behaviors of the solutions:
\ba\label{BIF314ap}
&\phi \sim t_{hop}r^{\alpha_t}-\dfrac{4  \delta\rho^2 q^4
     t_{hop}^3r^{
    {3} \alpha_t}}{(2\alpha_t+1) \alpha_t (q^2\delta\rho^2 +
      {3} \alpha_t +
      {9} \alpha_t^2)} + {\cal O}(r^{5\alpha_t}) + \varphi_{0} r^{-1-\alpha_t}(1 + \dots) , \nonumber \\
      &A_t^{(l)}\sim\mu +\rho_{(l)}r-(-1)^l\dfrac{\delta\rho q^2 r^{2\alpha_t+1} t_{hop}^2}{
 (2\alpha_t+1) \alpha_t} + {\cal O}(r^{4 \alpha_t + 1}),
\ea
where $\delta \rho=\rho_{(1)}-\rho_{(2)}$ and $\alpha_t=(-1+\sqrt{1-4q^2\delta\rho^2})/2$. We include the subleading corrections in the above asymptotic expansion because subleading corrections are important even for large $r$. For example, we find finitely many correction terms to the hopping term in the expansion of $\phi$ for a specific choice of $q^2 \delta\rho^2$. The situation in the gauge field expansion is the same. The condition to  stay above the BF bound, i.e. to keep real $\alpha_t$,  is
\be\label{BF2}
4 q^2 \delta \rho^2 \leq 1\,.
\ee
To keep real $\alpha_t$, namely, we restrict to the case $|\delta \rho |=1$ for the integer occupations $\rho_{(l)}$ and for our choice of charge \eqref{qchoice}.
Since at zero hopping the IR potential only contributes an additive shift to the free energy, we still identify $t_{hop}$ and  $\tilde{\varphi}\equiv \varphi_{0} (1-2\Delta_{\phi})$ with the hopping term and VEV of this operator in the non-homogeneous phase. The additional factor of $1-2\Delta_{\phi}(=-1/5)$ appears from the requirement of Ward identities in the field theory side~\cite{Freedman:1998tz,Klebanov:1999tb}. We still relate the imaginary part of VEV with the current of our theory as discussed in section 4. $\alpha_t$ encodes the anomalous dimension of this operator due to the interactions in the non-homogeneous phase.

\begin{figure}[htbp]
  \begin{center}
   \includegraphics[height=7cm]{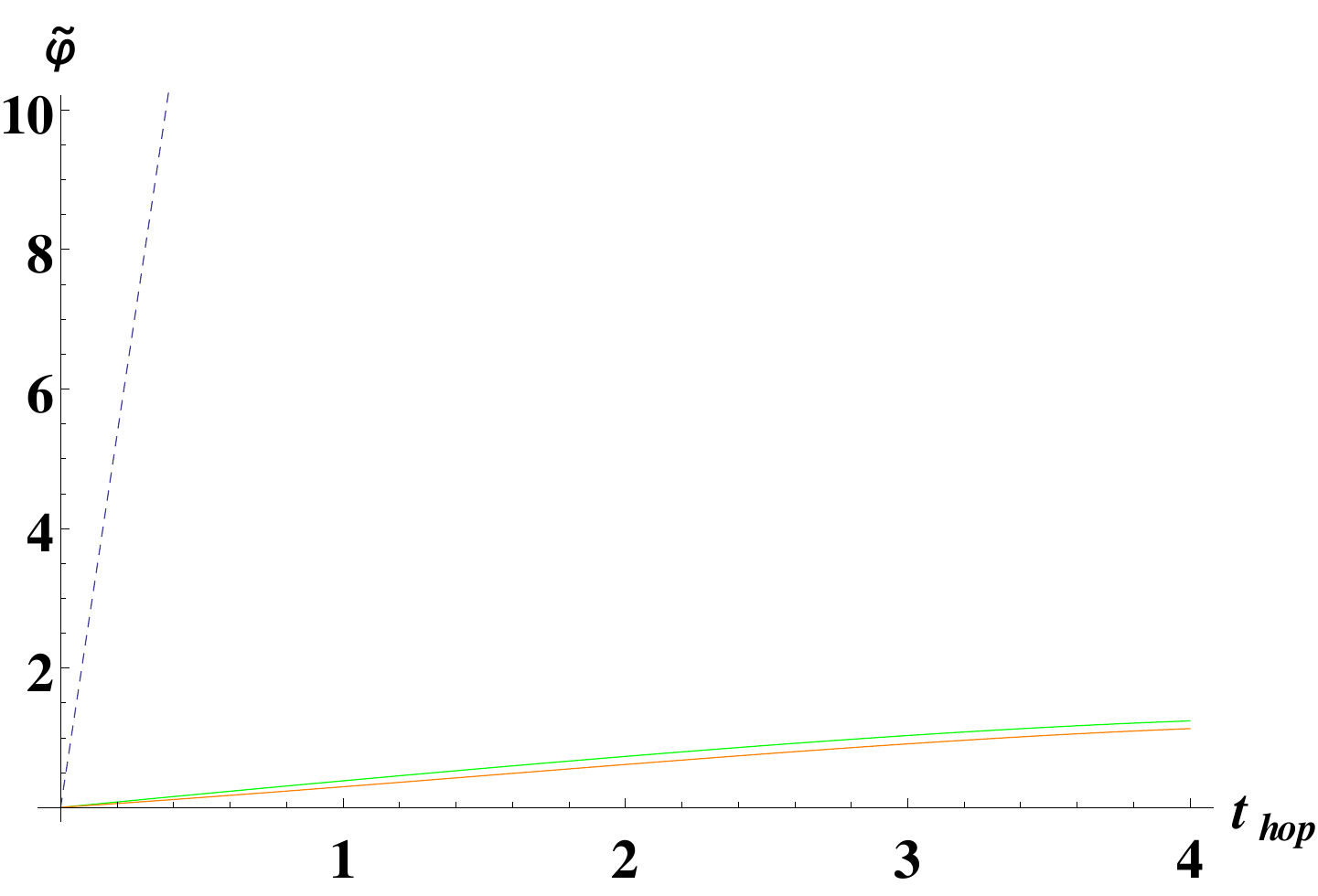}
  \caption{
  VEV $\tilde{\varphi}$ as the function of $t_{hop}$ ($\Lambda=1$) in the non-homogeneous phase for the mixed Neumann conditions.  Orange line: $w^{2}=0$. Green line: $w^{2}=1$. Dashed line: $\varphi^{Mott}$ in the Mott insulator phase for $w^{2}=1$. We see that the VEV in the inhomogeneous phase is always much smaller than the VEV in the Mott phase.
  }
    \label{fig:vphinonhom}
  \end{center}
\end{figure}

Our numerical procedure is as follows: instead of shooting from the IR to the UV we shoot from the UV to the IR, and vary the VEV  $\varphi_{0}$ until the IR boundary condition \eqref{IRB16ap} is satisfied. We plot the VEV $\tilde{\varphi}$ in the non-homogeneous phase as a function of $t_{hop}$ in Fig. \ref{fig:vphinonhom}. The VEV $\tilde{\varphi}$ increases as $w^{2}$ increases. As expected, the IR boundary condition changes with $w^{2}$, and hence the whole solution and in particular the UV VEV varies. The results in the non-homogenous phase should be compared with those in the Mott insulator phase for $w^{2}=1$ (Dashed line) - in this case the VEV is much larger. Since we expect a small VEV in the Mott phase, it would hence be preferable from a model building point of view to choose the vanishing VEV boundary conditions from the body of this paper in the Mott phase, and the mixed Neumann boundary conditions in the inhomogeneous phase.

\begin{figure}[htbp]
  \begin{center}
   \includegraphics[height=7cm]{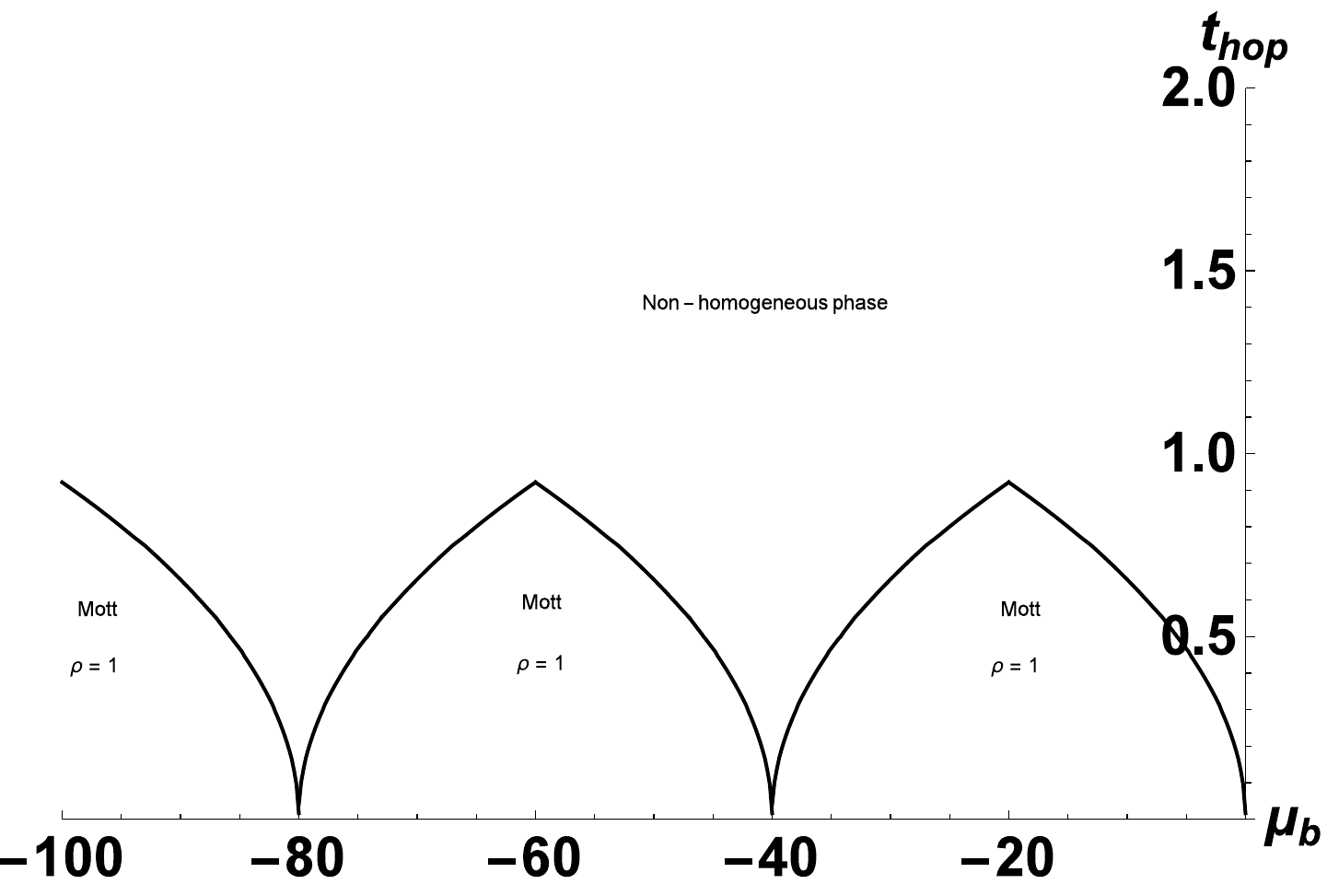}
  \caption{
  Phase structure of the two-site model for the mixed Neumann boundary conditions, for $r_h=40$, $w^2=1$, and $\Lambda =1$. Note that $\mu_{b}=\mu +U/2$. Inside the lobes, the charge density on both sites is equal, analoguous to the situation in~sec.~4. We identify this homogeneous phase with the Mott insulating phase. For large $t_{hop}$, there are regions where the non-homogeneous states are thermodynamically favored. The basic structure of the phase diagram is unchanged from the case discussed in the body of this paper.
}
    \label{fig:Lobeap}
  \end{center}
\end{figure}

\begin{figure}[htbp]
  \begin{center}
   \includegraphics[height=7cm]{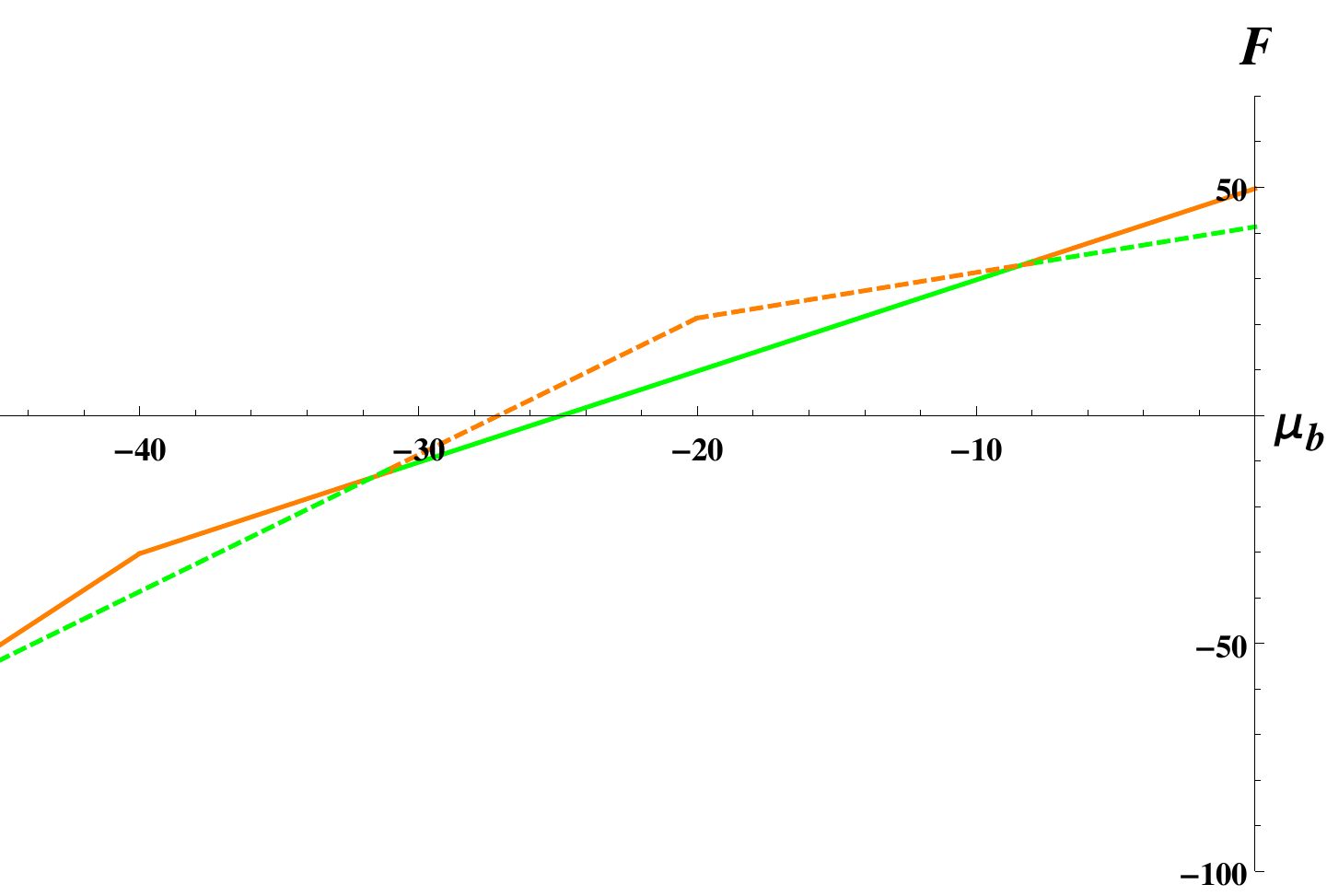}
  \caption{
  The free energy plotted as the function of $\mu_{b}$ for $t_{hop}=0.6$.  The color coding is defined as follows: Green lines show the thermodynamically favored phase, while red lines show unstable phases of higher free energy. Solid lines show the Mott phase, dashed lines show the inhomogeneous phase. The figure reflects that the Mott insulator phase is not thermodynamically favored between lobes of Fig. 12. Thus, first order level-changing phase transitions take place between homogeneous phase and non-homogeneous phases at finite hopping parameter.}
    \label{fig:Lobecheap}
  \end{center}
\end{figure}

\begin{figure}[htbp]
  \begin{center}
   \includegraphics[height=7cm]{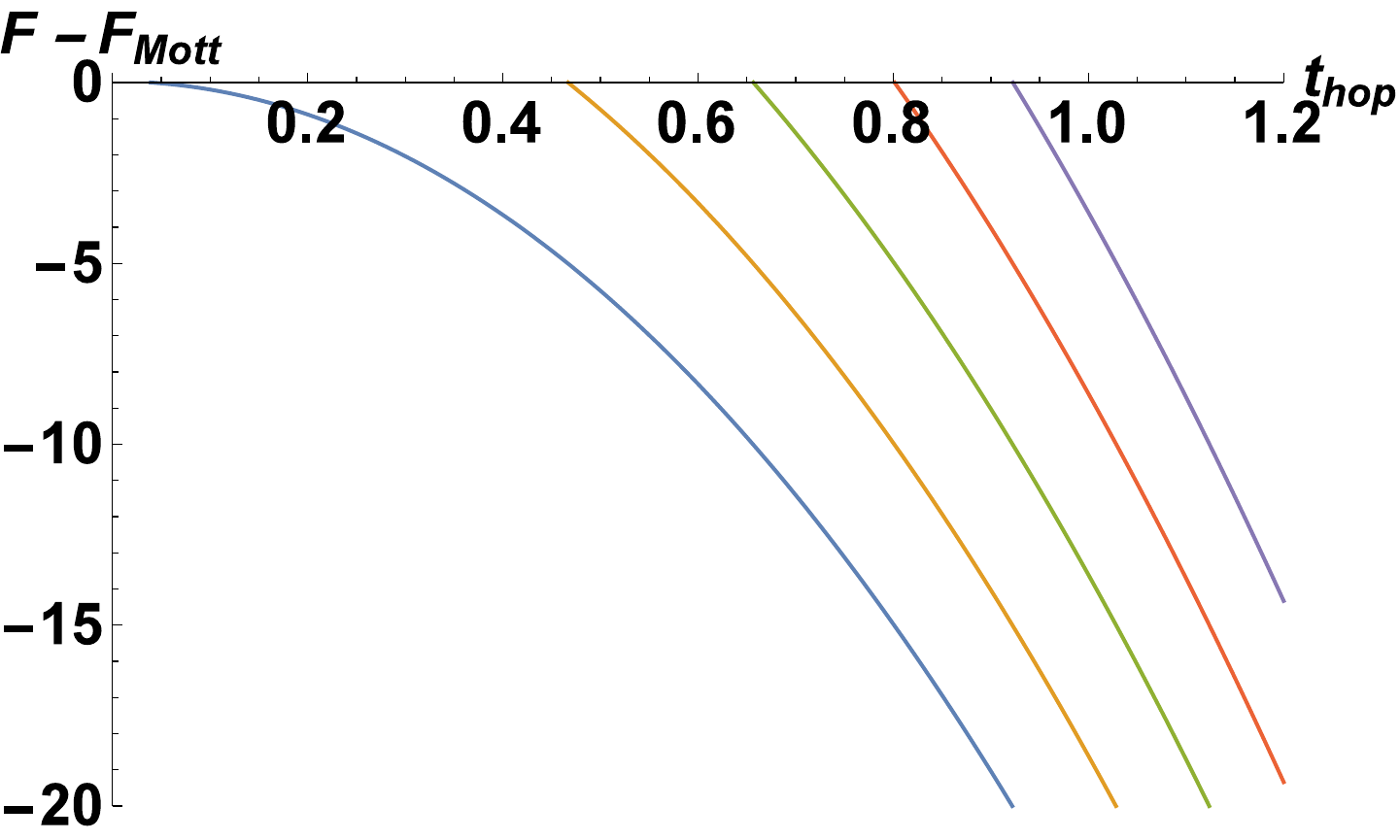}
  \caption{
  Difference of the free energy $F-F_{Mott}$ between the non-homogeneous phase and the Mott insulator phase $(\rho=2)$ as the function of $t_{hop}$. From the left to the right, the free energy is plotted for fixed $\mu_b/U=\mu/U +1/2=-1,-1.125,-1.25,-1.375,-1.5$, respectively. It implies that for the cusp point at $\mu_b/U=-1$, the non-homogeneous phase is always favored when $t_{hop}\neq 0$. $\mu_b/U=-1.5$ is a particle-hole symmetric point. After crossing this point, one attains the curves again in reversed order until one reaches the next cusp point at $\mu_b/U=-2$.
  }
    \label{fig:Lobew1freeap}
  \end{center}
\end{figure}

\noindent

The bi-fundamental's action is UV divergent in the non-homogeneous case $\delta\rho =\pm1$. To cancel this divergence,
the following counter-terms should be added :
\ba
S_{cut,2}=-\alpha_t\int_{r=R} dt\sqrt{-h}\phi^2.
\ea
The free energy is then given by the holographically renormalized action as
\ba
F=-(S_{kin}+S_{matter}+S_{cut}+S_{cut,2})/\beta .
\ea
Note that the diagonal gauge field $A_V=A^{(1)}+A^{(2)}$ decouples from the remaining parts. Thus, $F$ can be rewritten as
\ba\label{FRE315ap}
F=\mu \sum_i\rho_{(i)}+E\Big(\sum_i\rho_{(i)},\delta\rho, t_{hop}\Big).
\ea
The above formula shows that the energy $E(\sum_i\rho_{(i)},\delta\rho, t_{hop})$ is independent of the chemical potential at zero temperature at least.

The phase structure of the two-site model is plotted numerically in Fig. \ref{fig:Lobeap}
for $r_h=40$, $w^2=1$, and $\Lambda =1$. Note that the chemical potential  $\mu_b$ is defined in \eqref{MAT28}. The lobe-shaped phase structure of the Bose-Hubbard model \cite{Fisher:1989zza} is also realized with the mixed Neumann boundary conditions, c.f. Fig.~\ref{fig:Lobeap}. For  finite $t_{hop}$, there are regions where the inhomogeneous state is favored. Furthermore, the non-homogeneous phase extends to the $\mu$-axis at $\mu_b/U\equiv\mu/U +1/2=0,-1,-2$ as seen in Fig. \ref{fig:Lobeap}. A small VEV $\sim\epsilon \phi $ is expected near these critical points on the $\mu_{b}$-axis.

The amplitudes of all lobes can be changed by arranging the parameter $w$ in the IR potential accordingly. The amplitude is decreased as $w$ becomes large. Note that when $\mu$ decreases, the tips of lobes are not changed in our model even with the mixed Neumann boundary conditions Fig. \ref{fig:Lobeap}, while the tips of the lobes decrease as $1/\rho_{(1)}$ in the actual Bose-Hubbard model~\cite{Fisher:1989zza}.This shows that the height function of the lobes solely depends both on the choice of IR potential, as well as on the choice of boundary conditions.

The free energy is plotted as a function of $\mu_b$ for  fixed hopping parameter $t_{hop}=0.6$ in Fig. \ref{fig:Lobecheap}. There, the green lines are the thermodynamically prefered phases, while the red lines are the nonprefered phases with higher free energy, and the solid line corresponds to the Mott phase, while the dashed line corresponds to the inhomogeneous phase. The level-crossing phase transitions of first order are found between the homogeneous phase and non-homogeneous phase at finite hopping, in complete analogy to the free boundary conditions employed in the body of the paper. $F$ for the inhomogeneous phase is plotted as a function of $t_{hop}$ in Fig. \ref{fig:Lobew1freeap}. Again, a second order phase transition is found near  $t_{hop}=0$ in Fig. \ref{fig:Lobew1freeap}. However, a first order phase transition is found between the Mott phase and the non-homogeneous phase for all values of the hopping parameter, except for $t_{hop}=0$. This is also in accordance with the findings in the main part of this paper.  We conclude that we find no qualitative difference in the phase structure of our model between the boundary conditions we discuss here and the boundary conditions we use in the body of the paper.

%%%%%%%%%%%%%%%%%%%%%%%%%%%%%
%%%%%%%%%%%%%%%%%%%%%%%%%%%%%
%%%%%%%%%%%%%%%%%%%%%%%%%%%%%

\end{document}